\documentclass[12pt]{article}
\usepackage{latexsym}
\usepackage{epsfig,amssymb,euscript}
\usepackage{amsmath}
\numberwithin{equation}{section}

\def\be{\begin{equation}}
\def\ee{\end{equation}}
\def\bea{\begin{eqnarray}}
\def\eea{\end{eqnarray}}
\newcommand{\nn}{\nonumber}

\newcommand\R{\mathbb{R}}

\newcommand\cF{\mathcal{F}}
\def\bA{\mathbf{A}}
\def\bF{\mathbf{F}}

\def\bbF{\mathbb{F}}
\def\cA{\mathcal{A}}
\def\cF{\mathcal{F}}
\newcommand\cp{\mathbb{CP}}

\newcommand\diff{\mathrm{d}}
\newcommand{\ex}{\mathrm{e}}
\newcommand{\de}{\partial}
\newcommand{\vol}{\mathop{\mathrm{vol}}\nolimits}

\def\slash#1{\ooalign{{\text{$#1$}}\crcr \hss\big/\hss}}
\let\Box\square
\def\nref#1{(\ref{#1})}

\begin{document}
\begin{titlepage}
\begin{center}
\today

\vskip 2cm {\Large \bf
Comments on string theory backgrounds \\[3mm]
with non-relativistic conformal symmetry}

\vskip 1.5cm
{Juan Maldacena, Dario Martelli$^*$, and Yuji Tachikawa}
 \vskip 0.8 cm

{\em Institute for Advanced Study\\
1 Einstein Drive, Princeton, NJ 08540, U.S.A.}\\

\end{center}

\vskip 3cm

\begin{abstract}
\noindent
We consider non-relativistic
conformal quantum mechanical theories that arise by doing
discrete light cone quantization of field theories. If the field theory has a
gravity dual, then the conformal quantum mechanical theory can have a gravity dual
description in a suitable finite temperature and finite density regime.
Using this we  compute the thermodynamic properties of the system.
We give an explicit example where we display both the conformal quantum mechanical theory
as well as the gravity dual.
We also discuss the string theory embedding of certain
backgrounds with non-relativistic conformal symmetry that were recently discussed.
Using this, we construct finite temperature and finite density solutions,
 with asymptotic non-relativistic conformal symmetry.
In addition, we derive consistent Kaluza-Klein truncations of type IIB supergravity to a five
dimensional theory with massive vector fields.

\end{abstract}

\vfill
\hrule width 5cm
\vskip 5mm

{\noindent $^*$ {\small On leave from: \emph{Blackett Laboratory,
Imperial College, London SW7 2AZ, U.K.}}}

\end{titlepage}
\pagestyle{plain}
\setcounter{page}{1}
\newcounter{bean}
\baselineskip18pt


\tableofcontents

\section{Introduction}

Recently some gravity backgrounds
with non-relativistic conformal symmetry
were discussed \cite{son,mcgreevy,goldberger,barbonfuertes}\footnote{For earlier work on non-relativistic
conformal structures see e.g.  \cite{Horvathy:2003rb, Duval:1990hj} and references therein.}.
The idea is that such backgrounds would be gravity duals of conformal quantum mechanical systems,
which are useful for describing certain condensed matter systems (see \cite{son,mcgreevy,nishidason}
for references).

Here we make some comments on these constructions. These constructions involve performing
the discrete light cone quantization (DLCQ) of certain field theories and
their gravity duals. As it is well known,
the DLCQ  of a field
theory gives a non-relativistic system. Thus the DLCQ of a conformal field theory is expected
to give a conformal quantum mechanical
system. On the other hand, DLCQ quantization is subtle. In particular,
when we perform DLCQ of gravity backgrounds we cannot naively
apply the gravity approximation, since there is a circle becoming
very small. Fortunately, if one considers a sector with large
non-zero light-cone momentum  one can find regions in the geometry
where the circle has a non-zero size so that computations can be
trusted. In this paper we will discuss some backgrounds that arise
in string theory that display a non-relativistic conformal
symmetry. One particular example where we know the conformal
quantum mechanical theory and the corresponding gravity background
is the theory that arises when we do the DLCQ of the M5 brane
theory.
In this case the conformal  quantum mechanics was discussed in
\cite{ABS} and we will  review it below. We will discuss some
gravity backgrounds that can be used to perform computations which
can be trusted and are predictions for results in the conformal
quantum mechanical theory. Conformal quantum mechanical systems
were studied in the context of black hole physics and AdS$_2$, see
\cite{stromingerlectures} for a review and further references.

The backgrounds considered in \cite{son,mcgreevy} enjoy non-relativistic conformal
symmetry even before taking the DLCQ limit.
In this paper we embed these backgrounds in string theory   and discuss their deformation at
finite temperature and finite density.
In one particular case we can
relate these backgrounds to
certain non-commutative dipole theories studied in \cite{dipolelightlike}.
 There is a simple procedure that allows
us to introduce such non-commutativity both in the field theory and in gravity \cite{dipolegravity}.
 A non-relativistic
quantum mechanical system is expected to arise when we perform a DLCQ quantization of such a theory.
Due to the non-commutative interpretation of the background we find that certain quantities are
independent of the non-commutativity in the planar approximation, since non-commutativity does
not change certain planar diagrams \cite{filk}. Thus for many observables computations in
the backgrounds \cite{son,mcgreevy} are the same as in asymptotically AdS backgrounds with
 $x^-$ compactified.

We also discuss consistent type IIB Kaluza-Klein
reductions of AdS$_5 \times Y$ backgrounds, where $Y$ is a Sasaki-Einstein manifold,
 to five-dimensional systems
involving massive vector fields. Our motivation is
that these truncations admit solutions with (asymptotic)
non-relativistic conformal symmetries, of the type
discussed in \cite{son,mcgreevy}. Such Kaluza-Klein reductions
might be useful also for other
purposes.

The contents of the paper are organised as follows: we start in Sec.~\ref{DLCQ} by
discussing how we obtain a quantum mechanical system with non-relativistic conformal symmetry
via the discrete light cone quantisation of relativistic field theories. We also discuss subtleties inherent in the construction.  In Sec.~\ref{gravitybackgrounds} we construct the gravity dual of these systems,
which can be trusted when we put sufficient amount of momentum along the compactified dimension
so that it becomes a spacelike circle with large radius.
In Sec.~\ref{consistent} we study the consistent Kaluza-Klein reductions of type IIB supergravity
on $S^5$ or any other Sasaki-Einstein space which retain massive gauge fields, and we show
that these reductions can be used to construct some of the supergravity solutions discussed
in Sec.~\ref{gravitybackgrounds}. We conclude the paper with a short discussion in Sec.~\ref{conclusions}.
We have a few Appendices where the reader can find the details of the calculation.

\medskip

{\bf Note added}: the authors learned just before completion that
there will appear two papers, one by  A. Adams, K. Balasubramanian, and J. McGreevy \cite{ABM}
and another by C. P. Herzog, M. Rangamani  and S. F. Ross \cite{HRR}
which  have some overlap with our present paper.

{\bf Note added in v5}:  Appendix~\ref{powerlaw} was added on Nov.~2008 to explain the simple scaling behavior
of the thermodynamic quantities, which is a generic consequence of the gravity dual being a DLCQ of a higher dimensional theory.

\section{Non-relativistic theories from DLCQ}
\label{DLCQ}

\subsection{DLCQ of relativistic theories in Minkowski space}

It is well known that the light cone quantization of a relativistic theory looks
like a non-relativistic theory. Choosing light cone coordinates $x^\pm = t \pm x^3$ we
see that
 the mass shell condition for a massive particle looks like
\begin{equation}
-p_+ = { \vec p\,{}^2  \over (- 4 p_-) } +  { m^2 \over (- 4 p_- ) }
\end{equation}
which looks like the energy of a non-relativistic particle of mass $M \sim -p_-$ in a constant
potential.
We find it useful to write things in terms of $p_\pm$ with the lower index since that
is the momentum that is canonically conjugate to $x^\pm$ translations. One minor disadvantage
is that they are negative definite. Thus our $p_- = - p^+/2$ if we start with the ordinary
Minkowski metric, $\diff s^2 = - \diff x^+ \diff x^- + \diff \vec{x}^2$.

In a relativistic theory $p_-$ is a continuous variable.
 We can make it discrete by compactifying the light cone direction
$x^- \sim x^- + 2 \pi r^- $ \cite{maskawa}. We then find that $p_-$ is quantized as
\begin{equation}
-p_- = { N \over r^- }
\end{equation}
where $ N \geq 0$. This is called ``discrete light cone quantization''. Note that the
parameter $r^-$ can be changed by doing a boost in the $+-$ directions, which is a symmetry
of the relativistic theory. This boost is broken by the compactification. However, the fact
that it is a symmetry in the original theory
implies that theories with different values of $r^-$ are related by a simple rescaling
of the generators. Though in  the formulas below
we will keep $r^-$ explicitly, one could set $r^-=1$ without loss of generality.

If, in addition, the relativistic theory is conformal invariant, then this procedure would
formally lead to a conformal invariant quantum mechanical theory, with a symmetry group
which is called the ``Schr\"odinger group''\footnote{The name comes from the fact that it is the
dynamical symmetry group of the ordinary Schr\"odinger equation for a free particle.}.
This is simply the subset of the conformal generators
which commute with $p_-$. See Appendix \ref{conservationlaw} for some details.
This fact was already noted in \cite{ABS} where the DLCQ of the theory on M5 brane was studied.

As explained in \cite{shjp}, DLCQ is very subtle.
One has to be careful about the  zero modes.
In general these zero modes (modes with $p_-=0$)
 are described by an interacting theory which is
obtained by taking the original theory and placing it on a very small spatial circle of vanishing
size. Thus, one has to solve the problem of a field theory in one less dimension.
 For example, if we start with 3+1 dimensional ${\cal N}=4$ super Yang Mills, the zero
mode dynamics is described by a 2+1 dimensional conformal field theory which is the IR limit
of 2+1 dimensional super Yang Mills. This is also the theory that lives on M2 branes.
Thus, the proper analysis of the dynamics of a DLCQ theory is fairly non-trivial, but it can be
done in principle.
  This point should be kept in mind when we discuss various theories in this paper.
 Proposals for the DLCQ of
 ${\cal N}=4$ super Yang Mills were made in \cite{ganorsethi,kapustinsethi}.
We will not give a totally explicit description of the field theory side in this case,
leaving a complete analysis of this issue for the future. Note that in this case we get a family
of conformal quantum mechanical systems that arise by taking different expectation
values of $A_-$ (and the
dual photon) \cite{ganorsethi,kapustinsethi}.

The DLCQ procedure we outlined above is a way of generating examples of
conformal quantum mechanical systems. Writing out explicitly the quantum mechanical system
requires a proper  analysis of the zero modes.  This is an important issue for understanding
precisely the nature of the corresponding non-relativistic quantum mechanical system.
In particular, one would like to write down the Schr\"odinger equation for the quantum
mechanical system. We discuss one specific example below.

Finally, note that the discussion in \cite{stromingeradstwo} that links AdS$_3$ to AdS$_2$ and
a possible conformal quantum mechanical dual can be interpreted as a DLCQ quantization of
AdS$_3$ in a sector with nonzero $P_-$.

If the parent relativistic theory has  a gravity dual which is weakly coupled one can hope
to have a gravity description of the corresponding  conformal quantum mechanical system.
For example, if the parent theory is ${\cal N}=4$ super Yang Mills, then it also has a
dual description as a gravity (or string) theory on AdS$_5 \times S^5$ when the 't Hooft coupling
$g^2_{YM} k$ is large. We denote by $k$ the rank of the gauge group in order not to confuse it with
$N$, which is the number of quanta of
the light cone momentum. Thus, one would hope that by performing the DLCQ procedure on both sides
one would get a strongly coupled quantum mechanical system that is dual to a weakly
coupled gravity solution. The DLCQ limit of the gravity dual is simply given by
identifying the $x^-$ direction in the bulk \cite{barbonfuertes,goldberger}.
One should remember, though, that this identification
 is not as innocent as it looks. When we
periodically identify the $x^-$ direction in the bulk we are performing a drastic change
in the theory. For example, the correct graviton scattering amplitude in the DLCQ theory
and the naive one  (obtained by truncating tree level
 graviton scattering amplitudes in the theory before the
DLCQ) are not the same \cite{dinerajaraman}.

The situation changes for the better when we
 introduce a large amount of momentum $N \sim -p_- r^-$
in the
DLCQ direction. In fact, one is interested in the sector of the theory with non-zero values
of $N$. In the bulk, this has the effect of making the size of the $x^-$ circle spacelike in some
interesting regions of the geometry. Thus, for large enough $P_-$, or large enough $P_-$ density,
one can indeed use the gravity description
for computing certain properties of the system. We will describe this in more detail when
we talk about the gravity solutions. Let us first describe a specific example and
 some simple variations of this
construction.

\subsection{DLCQ description of the M5 brane theory}
\label{M5}

There is one case where a fairly explicit description of the DLCQ is available in the literature.
It is the case of the M5 brane theory, which is a 5+1 dimensional conformal field theory.
In DLCQ with $N$ units of momentum, this becomes a certain conformal quantum mechanics
theory constructed as follows \cite{ABS}.
We start with $k$ fivebranes and $N$ units of momentum
along a compactified spatial direction. The small radius limit,
which leads to the DLCQ description, forces us to
 perform a   duality to end up with $N$ D0 branes and $k$ D4 branes,
and to take the
low energy limit of this system.
The result is a quantum mechanics theory which is a sigma model on the Higgs branch of a certain
theory with 8 supercharges. It is the quantum mechanics on the moduli space of $N$-instantons.

This system has four-dimensional Galilean invariance and   conformal invariance, so
it has the Schr\"odinger symmetry, but it is quite unlike the non-relativistic conformal system
which is discussed recently in the literature, e.g.~$N$ fermions interacting via contact interaction.
Namely, each of the $N$ instantons is not pointlike
but has a size parameter which is affected by the dilatation;
there is no obvious second-quantized framework, and so on.

The resulting quantum mechanics is the following. We start with
 a $U(N)$ gauge theory with an adjoint hypermultiplet and
$k$ fundamental hypermultiplets. The bosonic variables involve two complex adjoint matrices
$X, ~\tilde X$ and $k$  complex scalars $q_i$ in the ${\bf N}$ of $U(N)$ and $k$ scalars
$\tilde q^i$ in the $\bar {\bf N}$ of $U(N)$. These fields are constrained by
\be \label{condit}
[X, X^\dagger]- [ \tilde X, \tilde X^\dagger] + q_ i q_i^\dagger - ({ \tilde q}^i)^\dagger ( \tilde q^i) =0 ~,
~~~~~~~~~~~~ [ X , \tilde X] + q_i \tilde q^i =0.
\ee
and we quotient by $U(N)$ gauge transformations.
This gives a space of $4 N k$ real dimensions which is a hyperk\"ahler manifold. The metric
is the induced metric in the ambient space. It can be found as follows: one
constructs the K\"ahler potential of the ambient space
$C = | q_i|^2 + | { \tilde q}^i |^2 + |X|^2 + | \tilde X |^2 $, which also gives
the K\"ahler potential of the metric in the moduli space, after choosing
complex coordinates for the moduli. $C$ is also the special conformal generator in the
 quantum mechanical theory.  This defines a conformal quantum mechanics,
 which becomes superconformal once we include also the fermionic degrees of freedom \cite{ABS}.
For further details on the definition of the quantum mechanical theory and its symmetries see
\cite{ABS}. See also \cite{stromingerlectures} for a nice review on conformal and
superconformal quantum mechanics.

\subsection{DLCQ of  a conformal field theory on a plane wave background}
\label{planewave}

A small variation of the preceding theme is to start with a relativistic theory on a plane wave background
\begin{equation}
\diff s^2 \,=\, - \diff x^+ \diff x^-  - \vec x\,{}^2 (\diff x^+)^2  + \diff {\vec x}\,{}^2
\label{ppwave}
\end{equation} where $\vec x$ stands for the transverse spatial directions.
For a general field theory this
also gives us a quantum mechanical system now with particles in a harmonic oscillator
potential. The Hamiltonian is then
\be
 - p_+ \,=\, { \vec p\,{}^2 \over (-4 p_-) } +  (-p_-) \vec x\,{}^2~.
 \ee
The isometries of the plane wave are symmetries acting on the quantum mechanical
theory.
If the field theory is conformal we have further symmetries and the resulting theory is
precisely the same as the one we obtained when we started from flat space but with a different
choice of Hamiltonian  \cite{flattopp}.
This can be understood considering the $SL(2,\R)$ subgroup of the Schr\"odringer group
which includes the Hamiltonian $H$ the dilatation $D$ and the special conformal transformation
$C$. Then the Hamiltonian on the plane wave background is
\bea
H_{osc} \,=\, L_0 \,= \, { 1 \over 2} ( H + C)~.
\eea

Another variation of the idea is to take the following form of the plane wave metric
when the transverse direction is two-dimensional (or more generally even dimensional):
\begin{equation} \label{magfield}
\diff s^2 \,=\, - \diff x^+ (\diff x^-  - 2\rho^2 \diff\hat \psi ) + \diff\rho^2+\rho^2\diff\hat \psi^2~,
\end{equation}
where we took $\vec x =(\rho\cos \psi,\rho\sin \psi)$ and chose $\hat \psi=\psi-x^+$.
A particle with fixed $p_-$ moving on this metric reduces to a non-relativistic particle moving
in the transverse space in the presence of a constant magnetic field
  $\diff(\rho^2\diff\psi)$ with no potential.
The new Hamiltonian is related to the one above by
\begin{equation} \label{magoscil}
H_{mag}\,=\, H_{osc}-J~,
\end{equation}
where $J$ is the angular momentum associated to the rotation in the $\vec{x}$ plane.

\subsection{DLCQ of dipole theories}

Another variation is to consider a certain non-commutative theory, called a ``dipole theory''
\cite{dipolelightlike} (see also \cite{dipolegravity,dipoleone,dipoletwo}).
This is a theory were the field multiplication is defined via a star product. In order to
define the star product we use the conserved charge $p_-$ and also another global symmetry
charge, $Q$ of the system.
The star product is then defined as follows
\begin{equation} \label{starproduct}
f * g \,=\, \ex^{ i 2 \pi \sigma ( P_-^f Q^g - P_-^g Q^f )  } f g~,
\end{equation}
where $fg$ is the ordinary product and $\sigma$ is an arbitrary parameter. $(P_-^f,Q^f)$ and
$(P^g_-,Q^g)$ are the values of $P_-$ and $Q$ for $f$ and $g$ respectively.
 We are imagining that $f$ and $g$ have well defined values
for both charges $P_-$ and $Q$, and we can get the product for more general functions of $f$ and $g$
by the ordinary distributive property of the product.

If the original theory has a symmetry that commutes with the two charges that appear in the definition
of the star product \nref{starproduct}, then it will also be a symmetry of the theory after the
star product deformation. The generators of the conformal group that commute with $P_-$ are
the Schr\"odinger subgroup. Thus we have the Schr\"odinger symmetry even before
compactifying the coordinate $x^-$. If we also compactify $x^-$ and perform a DLCQ quantization
we expect to get a non-relativistic conformal system. Our reason for introducing
these exotic theories is that, in some cases,
 their gravity duals, derived in \cite{dipolelightlike,dipolegravity},
 are given by the metrics introduced in
 \cite{son,mcgreevy}. For more details on these theories
  see \cite{dipolelightlike,dipolegravity,dipoleone,dipoletwo} .

\section{DLCQ of string or M theory  backgrounds }
\label{gravitybackgrounds}
\subsection{DLCQ of AdS}\label{DLCQofAdS}

As we saw in the previous section, one can construct  a system with Schr\"odinger symmetry
 by taking the DLCQ of a relativistic conformal theory.
  If the relativistic CFT   is dual to AdS we should then
   consider the DLCQ of AdS space.
We can write the AdS metric in the Poincar\'e patch as
\begin{equation}
\label{adsmetr}
\diff s^2 \,= \, {\diff r^2 \over r^2 }+ r^2\left( - \diff x^+ \diff x^- + \diff \vec x\,{}^2\right)
\end{equation}
and take $x^- \sim x^- + 2 \pi r^- $. This fact has been noticed also in the recent papers \cite{barbonfuertes,goldberger}.

Since we are taking a circle to have zero size, we cannot  trust this geometry to
do computations.
 A similar  situation arises when we
consider the DLCQ of eleven dimensional supergravity. In that case the conjectured
correct description is in terms of a quantum mechanical theory given by
 $N$ D0 branes \cite{SUSSKIND} (see also \cite{BFSS}).
  This description is not the same as the one we get
 by taking the naive DLCQ of the gravity theory. For example, the scattering amplitude of three gravitons to three gravitons  in the naive supergravity approximation
 gives a different answer  than in the
 matrix model \cite{dinerajaraman}. In fact, when one defines the DLCQ limit carefully, as a limit of the theory
 on a very small spatial circle, one finds that the correct answer is given by the
 D0 matrix quantum mechanics and not by supergravity \cite{Seibergmm}.\footnote{Of course,
  the BFSS conjecture
 \cite{BFSS} is the idea that a suitable large $N$ limit allows one to recover the results with no compactification
 of the light like circle.  }

 Fortunately, not all is lost. Of course, what we really want to do is to put $N$ units of
 momentum on this space. When $N$ is small, the dynamics cannot be computed in terms of
 particles moving in the metric (\ref{adsmetr}), for reasons we have explained. Notice that this is
 true even when the radius of  AdS is large. In fact, the correct description of
 $N$ units of momentum in the DLCQ of  type IIB string theory in flat space is in terms
 of the field theory that lives on $N$ M2 branes on $T^2$ \cite{IIBDLCQ}.

  However, if we put a large amount of momentum, then the backreaction implies that
  the size of $x^-$ will be non-zero in some regions and we will be able to trust the metric.
  More concretely, we can consider the black three brane metric describing a finite
  temperature system
\bea
\diff s^2 \,= \,     \frac{1}{1- \frac{r_0^4}{ r^4}}  \frac{\diff r^2}{r^2} +  r^2 \left[ -\diff x^+ \diff x^-
+  \frac{ r_0^4}{4r^4} \left(\lambda^{-1} \diff x^+ + \lambda\diff x^- \right)^2   + \diff \vec x\,{}^2  \right]
   ~.
 \label{blackthree}
\eea
This is simply a boosted version of the ordinary black brane metric,
in the near-horizon region.
We can compute the density of $-P_-$, which we interpret as the ``particle density'', and
the energy density. We obtain (see Appendix \ref{thermodetails})
\begin{equation} \label{pplusminus}
 { N \over V_2} \,=\,  { r^- (-P_-) \over V_2 } \,  = \, {R_{AdS}^3 \over   G_N^5 } {    (r^-)^2 \lambda^2 r_0^4 \over 8 } ~,~~~~~~~
  { H \over V_2} \,= \, { (-P_+) \over V_2}  \,= \,  {R_{AdS}^3 \over   G_N^5 } { r_0^4 \over 16}    r^-~,
\end{equation}
where $V_2$ is the volume of the two spatial dimensions.

We can see that by tuning $\lambda$ and $r_0$ we get different values of particle density
  as well as energy density. We can also write this in terms of the temperature and
  chemical potential of the non-relativistic system
\begin{equation}
\label{temperature}
   { 1 \over T} \,=\, {   \pi \lambda \over r_0}  ~,\qquad
    {\mu_N  \over T}  \,=\, {   \pi  \over r_0 r^- \lambda}~.
\end{equation}
The entropy is given by
 \be \label{Entropy}
   S\, =\, {R_{AdS}^3 \over 4 G_N^5 } \lambda  ( 2 \pi r^-) {  r_0^3 \over 2 }  V_2~.
 \ee
and these thermodynamic quantities satisfy the first law
 \begin{equation}
\delta H\,= \, T \delta S-\mu_N \delta N~.
\end{equation}

The physical value of the radius of the $x^-$ circle is
\begin{equation}
{ R^-_{phys} \over l_s}\, =\, { R_{AdS} \over l_s}  { r_0  \over r} { \lambda r^-  r_0 \over 2 }~,
\end{equation}
therefore we can trust the gravity description as long as ${ R^-_{phys} \over l_s}  \gg 1 $. When it
   becomes smaller than one we might be able to do a T-duality and use an alternative description
   also. Notice that this size becomes small as we approach the boundary of AdS, $r \to \infty$.
   Thus, we cannot trust the metric near the boundary. It might be
    possible that after performing
   suitable U-dualities we might find a metric we can trust.

   Of course this solution implies that the thermodynamic properties can be simply extracted from
   the thermodynamic properties of the ordinary black D3 brane. We are only  looking at the
   same system in light cone gauge, so we just need to translate all quantities to light cone
   gauge.

\subsection{DLCQ of AdS$_5$ with a plane wave boundary: harmonic potential}

In this subsection we consider the gravity dual of the field theory with plane wave
boundary conditions, so that it is the gravity dual to the field theory on the
plane wave. Since the plane wave metric is conformal to flat space the metric is simply AdS, which can
be sliced in a way that make the plane wave at the boundary more manifest.
One way to find the slicing is to start with $\R \times S^3$ and take the Penrose limit
for a particle moving with large angular momentum in one of the angles of $S^3$.
Namely, we write
the AdS metric in global coordinates,
 \begin{equation}
 \diff s^2 \,=\, - (1+r^2) \diff t^2+ { \diff r^2 \over 1+ r^2} + r^2 \diff s^2(S^3) ~,
 \end{equation}
 which has $\R \times S^3$  as a  boundary.
Here  $ \diff s^2({S^3}) = \diff\theta^2+ \cos^2 \theta \diff\varphi^2 +
\sin^2 \theta \diff\psi^2 $.
We then define
\be
 x^+ \,= \, t   ~,~~~~~~~~ {x^- \over 2 R^2 }\, =\, t - \varphi ~,~~~~~ \theta \,= \,\frac{\rho}{R}
 ~,~~~~~~~~ r = R y  \label{RxS3planewave}
 \ee
 and we take the  $R \to \infty $ limit, keeping $x^\pm$ and $\rho$, $y$ fixed.
 We find the metric
 \be
 \label{adspp}
 \diff s^2\, =\, { \diff y^2 \over y^2 } + y^2 \left( - \diff x^+ \diff x^- - \rho^2(\diff x^+)^2 + \diff\rho^2 + \rho^2 \diff\psi^2 \right) - ( \diff x^+)^2~.
 \ee
Notice that  it is not necessary to take this limit to obtain (\ref{adspp}), and  we can also
 obtain it directly by writing AdS space  in the appropriate
 coordinates\footnote{In fact, \eqref{adspp} can be transformed into the metric recently discussed
 in \cite{barbonfuertes}  (\emph{cf}. equation (35) in that reference) by
 an obvious change of coordinates.
 }
  (see also \cite{goldberger}).
  Explicitly,  after the further change of coordinates
 \bea
 \sinh r \, = \, \rho y ~,\qquad z \,= \, \rho^2+ \frac{1}{y^2}~,
 \eea
the metric  \nref{adspp} becomes
\bea
 \label{adspp:ads}
 \diff s^2\, =\, \sinh^2 r  \, \diff \psi^2 +  \diff r^2 + \cosh^2 r
  \left[ - (\diff x^+ + \frac{\diff x^-}{2z})^2
  + \frac{(\diff x^-)^2 + \diff z^2}{4z^2}  \right]~.
 \eea
 This is AdS$_5$ sliced by AdS$_3$, which is
 the metric in square brackets.
 However, obtaining the metric \nref{adspp} as a limit
 will be useful later.
 The metric in (\ref{adspp}) cannot be trusted if $x^-$ is
 compact, since $x^-$ is a null direction.

We would now like to add some $x^-$ momentum and also raise the temperature of the system so
that  we have a black hole in a space which is asymptotic to \nref{adspp}. We obtain this black hole
 by starting with the five-dimensional Kerr-AdS black hole \cite{Hawking:1998kw}
 (see also \cite{AdSKerrFirstLaw})
 and performing a limit similar to the limit we performed above \nref{RxS3planewave}.
 We describe
 this in detail in appendix \ref{BHplane}. The final metric can be
 written as
  \bea \label{bhpw}
   \diff s^2 \,= \, { ( r^2 + \sin^2 \theta) \diff r^2
  \over (1 + r^2)^2  - 2 m } - ( 1 + r^2 \sin^2\theta ) ( \diff x^+)^2  -   \lambda (1 + r^2)
  \cos^2 \theta \diff x^+ \diff x^-
  \qquad \nn \\[2.8mm]
  + { ( r^2 + \sin^2 \theta) \diff \theta^2 \over \cos^2 \theta } + r^2\sin^2 \theta  \diff \psi^2
    + m { ( -2  \diff x^+ + (   \diff x^+ - \lambda\diff x^-) \cos^2 \theta)^2 \over 2
     ( r^2 + \sin^2 \theta )}~.
  \eea
  This metric depends only on one non-trivial parameter $m$. The parameter $\lambda$ can be
 absorbed into the redefinition of $x^-$ but it is convenient to keep it because it represents the amount of boost one performs when we take the limit.
We can bring this metric to a form which asymptotes to \eqref{adspp}
via the coordinate change:
\bea
\lambda^{-1} y^2 &= &(r^2+1)\cos^2\theta~,\nn\\
\rho^2y^2 &= &r^2\sin^2\theta~. \label{coc}
\eea
This is analogous to the coordinate change employed in \cite{Hawking:1998kw} to display the
AdS asymptotics of the Kerr-AdS black hole.
The result does not have a simple analytic form.
As an expansion in the new radial variable $y$, it is given by
\begin{align}
\diff s^2 & =  \,\left(1 - \frac{2 m}{(1+\lambda \rho^2)^2}{\lambda^2\over y^4}  \right)^{-1}
{ \diff y^2 \over y^2}
+y^2 \left( - \diff x^+ \diff x^- - \rho^2(\diff x^+)^2 + \diff\rho^2 + \rho^2 \diff\psi^2 \right)\nn\\[2mm]
&  \quad  - ( \diff x^+)^2 +    \frac{\lambda m}{y^2}\frac{\left((1+2\lambda\rho^2)\diff x^+ + \lambda\diff x^-\right)^2}{2(1+\lambda\rho^2)^3} + {\cal O}(y^{-4})
  \label{ppBH}
\end{align}
where we kept only terms of order lower than  ${\cal O} (y^{-4})$ with
 respect to the $m=0$ solution.

The $P_-$ and $P_+$ can be calculated by starting from the
expressions for the energy and the angular momentum in
\cite{AdSKerrFirstLaw} (also see \cite{Bhattacharyya:2007vs})
and taking the appropriate limit, see appendix \ref{BHplane}.
We find
\bea
\label{ener}
H\,=\,-P_+ &=&  { R_{AdS}^3 \over   G_N^5 }(2\pi
r^- ) \lambda { (1 + r_H^2 )^2 \over 16 } ~,~~~~~~~~
\\
N\,=\, -P_- r^- & =& { R_{AdS}^3 \over   G_N^5 } (2\pi  r^- ) r^- \lambda^2 { (1 + r_H^2 )^2 \over 32 }~ .
\label{partdens}
\eea
Recall that the Hamiltonian here is the oscillator Hamiltonian $H_{osc}$.

The horizon radius $r_H$ is given by the solution of $(r_H^2+1)^2-2m=0$.
As $r_H\to 0$ the metric becomes singular. For $r_H=0$ (or $2m =1$)
the energy and particle densities from \nref{ener},
\nref{partdens} are non-zero. For smaller densities or energies this black hole would not
be a good description. However, for $m=0$ the metric \nref{bhpw} does go to the plane wave
metric \nref{adspp}\footnote{  This is reminiscent to what happens for the BTZ black hole. }.

The Killing vector degenerating at the horizon is
\begin{equation}
\partial_+ + \frac1\lambda\frac{r_H^2-1}{r_H^2+1} \partial_-~.
\end{equation}
 Then the temperature and the chemical potential are \begin{equation}
T\,=\,\frac{r_H}{\pi}~,\qquad
\mu_N\,=\, \frac{1}{ r^- \lambda} \frac{r_H^2-1}{r_H^2+1}~.
 \label{tempmu}
\end{equation}
 The entropy of the system is
 \be
 S \, =\,  { R_{AdS}^3 \over 4 G_N^5 } (2\pi r^-) { \pi \over 2 }   \lambda r_H(r_H^2+1)~,
 \ee
and one can check again the first law
\begin{equation}
\delta H \,=\, T \delta S -\mu_N \delta N
\end{equation} is satisfied, as it should be.
We can also compute the size of the circle $x^-$ at the horizon which is
 \be
 { R^-_{phys} \over l_p} = { R_{AdS} \over l_p} r^- \lambda { ( r_H^2 +1) \cos^2 \theta \over
 2 \sqrt{ r_H^2 + \sin^2 \theta } }~.
 \ee
This becomes zero when $\cos\theta\to 0$. In fact, the circle $x^-$ shrinks
at $\theta=\pi/2$ at any value of $r$ as is evident from \eqref{ppBH}.
In the  coordinates $\rho,y$ introduced in \eqref{coc} these points correspond to
the limit $y\to 0$, $\rho\to\infty$. Since we do not have a large contribution to
the thermodynamic quantities from this region we assume that we can ignore this part of
the metric.

\subsection{DLCQ of AdS$_5$ with a plane wave boundary: magnetic field}

Here we briefly discuss the gravity dual of the field theory with plane wave
boundary conditions in the coordinates \nref{magfield} which lead to a  non relativistic
particle in a magnetic field when we fix $p_-$.
The Hamiltonian of this system is obtained shifting the oscillator Hamiltonian with the angular
momentum on the plane, as   in \ref{magoscil}. The metric is obtained by
performing a limit similar to the one
discussed in the previous subsection, starting with the two-parameter Kerr-AdS black hole
\cite{Hawking:1998kw}. This is discussed in Appendix \ref{magnetic} and here we present the final result
\begin{multline}
\diff s^2 \, =\,  \left(1-  \frac{2m\lambda^4}{y^4} + \frac{2m\lambda^6}{y^6}\right)^{-1}
\frac{\diff y^2}{y^2}+
y^2 \bigg[ - \diff x^+(\diff x^-  - 2\rho^2 \diff \hat\psi ) + (\diff\rho^2 +\rho^2\diff\hat\psi^2)
 \bigg] \\[2mm]
-(\diff x^+)^2 +\frac{m\lambda^6}{ 2y^2 } \left[\diff x^-+ \frac{\diff x^+}{\lambda^2} - 2\rho^2\diff \hat\psi  \right]^2~.
\label{maglimit}
\end{multline}

The temperature and the chemical potential are given by
\bea
T \, =\, \frac1{2\pi} \frac{2y_H^2-3\lambda^2}{\lambda\sqrt{y_H^2-\lambda^2}}~, \qquad
\mu_N = \frac{y_H^2-2\lambda^2} {r^- y_H^2\lambda^2}~,
\eea
and the entropy is
\bea
S \, =\, \frac{R^3_{AdS}}{4G_N^5}  (2\pi r^-) V_2 \frac12 \frac{y_H^4\lambda}{\sqrt{y_H^2-\lambda^2}}~,
\eea
where  the horizon is at $y=y_H$ with
\bea
1 -  \frac{2m\lambda^4}{y_H^4} + \frac{2m\lambda^4}{y_H^6} \, =\, 0~.
\eea
The $P_\pm$ densities are constant (see Appendix \ref{magnetic}), thus  the charges are
infinite on the $\vec{x}$ plane, and proportional
to the volume $V_2$, as is in the case of the translation  invariant black three brane metric.
In fact, after Kaluza-Klein reducing along $x^-$ we see that we get a translation invariant metric with
a constant Kaluza-Klein magnetic field.
The particle number and energy per unit volume are given by
\bea
\frac{N}{V_2} \,=\, \frac{R^3_{AdS}}{4G_N^5}  r^-  m\lambda^6~,\qquad
\frac{H}{V_2}\,= \, \frac{R^3_{AdS}}{4G_N^5}  r^-  \frac{m\lambda^4}{2}~,
\eea
respectively.  Recall that the Hamiltonian here is given by  $H_{mag}= H_{osc}-J$.
This implies that we can view this system as the system on with the Harmonic oscillator hamiltonian
but at a critical value of the chemical potential for the spin in the transverse plane.
One could find a metric interpolating between \nref{bhpw} and \nref{maglimit} by adding a general
chemical potential for the spin. Such metrics should correspond to more general limits of the
rotating Kerr black hole \cite{Hawking:1998kw}.

\subsection{DLCQ of AdS$_7$ with a plane wave boundary }
\label{ADSseven}

In this subsection we repeat the previous discussion for the AdS$_7$ case.
We write the $S^5$ as $\diff s^2(S^5) = \diff \theta^2  +
\cos^2 \theta \diff \varphi^2 + \sin^2 \theta \diff\Omega_3^2 $. We now start with
AdS$_7$ in coordinates where the boundary is $\R \times S^5$. After performing
the rescaling \nref{RxS3planewave} we get a metric similar to \nref{adspp} except
that $\diff \psi^2 \to \diff \Omega_3^2$. Starting from the  seven dimensional
Kerr-AdS  black hole and performing the same limit \nref{RxS3planewave}
we get a metric very similar to \nref{bhpw} except for $m \to m/r^2$ and $\diff \psi^2 \to
\diff \Omega_3^2$,
 \bea
 \label{bhpwseven}
   \diff s^2 \,= \, { ( r^2 + \sin^2 \theta) \diff r^2
  \over (1 + r^2)^2  -  { 2 m \over r^2 } } -
  ( 1 + r^2 \sin^2\theta ) ( \diff x^+)^2  -   \lambda (1 + r^2) \cos^2 \theta \diff x^+ \diff x^-
  \qquad \nn \\[2.8mm]
  + { ( r^2 + \sin^2 \theta) \diff \theta^2 \over \cos^2 \theta } + r^2\sin^2 \theta  \diff  \Omega_3^2
    + { m \over r^2 } { ( -2  \diff x^+ + (   \diff x^+ - \lambda\diff x^-) \cos^2 \theta)^2 \over 2
     ( r^2 + \sin^2 \theta )}~.
  \eea

We can also compute the entropy as a function of the temperature
and the chemical potential. The expressions for the temperature
and chemical potential are
\begin{equation}
T\,=\, \frac{1}{2 \pi} ( { 3 r_H  } + { 1 \over   r_H }),\qquad
\mu_N\,=\, \frac{1}{ r^- \lambda} \frac{r_H^2-1}{r_H^2+1}~,
\label{tempmusev}
\end{equation}
 where now the horizon radius obeys   the equation
 $ ( r_H^2 + 1)^2 - { 2 m / r_H^2 } =0$.
 The values of the energy and the momentum
can be uniquely fixed by first noting that  $E\propto \lambda m$ and $N\propto \lambda^2 m$
from the asymptotic form of the metric,
and second by demanding that the first law is satisfied. We obtain
  \bea
H\,=\,-P_+ &=&  { R_{AdS}^5 \over   G_N^7} (2\pi  r^-)\lambda { \pi\over 16 }  { r_H^2 (1 + r_H^2 )^2  } ~,~~~~~~~~
\\
N\,=\, - P_- r^-  &=&
{ R_{AdS}^5 \over   G_N^7} (2\pi r^- ) r^-   \lambda^2 { \pi  \over 64}  { r_H^2 (1 + r_H^2 )^2   }~ .
\eea
 The entropy is
 \be
 S \,=\, { R_{AdS}^5 \over 4 G_N^7 } ( 2 \pi r^-) \lambda { \pi^2  \over 4 } r_H^3 ( 1 +  r_H^2 )~.
 \ee
 In this case we see that as $r_H \to 0$ the energy
and the particle number both go to zero. On the other hand, we see that the temperature
in \nref{tempmusev} has a minimum. This suggests that the thermal ensemble will display
a phase transition similar to the Hawking--Page  we see when we treat AdS in global
coordinates \cite{HP,wittenthermal}. In fact, for a given temperature and chemical potential there
is another solution which is simply thermal AdS space with a gas of particles (we can only trust this
last solution if   we take $x^-$ to be non-compact). Comparing their free energies we find a phase
transition at $r_H=1$. Of course, the black hole is the stable solution for  $r_H>1$.
 We can evaluate this quantities in the particular case of AdS$_7 \times S^4$ background of M
 theory. In that case we find that
 \be
 { R_{AdS}^5  \over 4 G_N^7 } = { R^5_{AdS} R^4_{S^4} V_{S^4} \over 4 G_N^{11} }  = { 4 k^3
 \over    \, 3 \pi^2 } ~,
 ~~~~~~~ { R_{AdS} \over l_p} = 2 { R_{S^4} \over l_p } = 2 ( \pi k)^{1/3}~,
 \ee
 where $G_{N}^{11} = 16 \pi^7 l_p^9$ and $k$ is the number of M5 branes.
As in the five-dimensional case, the $x^-$ circle shrinks when $\theta=\pi/2$.
For moderate temperatures, with $r_H$ of order one, we can trust the M-theory background near
the horizon in the regime
$k\gg 1 $ and $N^3/k^7 \gg 1 $. By reducing to type IIA we can extend the region of validity
of the above formulas to $k\gg1 $, $N/k \gg 1$. The existence of the IIA string theory
dual suggests that this conformal quantum mechanics theory has an interesting 't Hooft limit.

 These quantities should describe the thermodynamic properties of the conformal matrix
 quantum mechanics described in \cite{ABS} and reviewed in Sec.~\ref{M5}.
 That quantum  mechanics theory is characterized by only two discrete parameters $k$ and $N$. Here
 $k$ is the number of fivebranes and $N$ is the amount of momentum. When we consider
 the Hamiltonian $H_{osc}= \frac12(H + C)$ we expect that the temperature would also be a non-trivial
 parameter since there is an energy gap of order one that is given by the confining
 potential. In the above metric the temperature translates directly into $r_H$. Finally,
 we see that the combination $r^- \lambda$, at fixed temperature, fixes the chemical potential
 for the light cone momentum $N$. Thus the metrics we have depend on the right number of
 parameters. Here $r^-$ is a trivial parameter and can be set to one without loss of generality.

\subsection{DLCQ of the dipole theory and its gravity dual}
\label{dipole:bh}

In this section we describe in more detail a particular example of
the dipole theories introduced in
\cite{dipolelightlike,dipoleone,dipoletwo}. As we explained above
these theories are based on the non-commutative $*$ product in
\nref{starproduct}. As a particular example we can consider
starting with ${\cal N}=4$ super Yang Mills in 3+1 dimensions.
This theory has an $SO(6)$ R symmetry that rotates six scalars,
$\phi_i$,  and their fermionic partners. We can consider the
particular $U(1)$ symmetry that rotates all pairs of scalars.
Defining  $W_j = \phi_{j} + i \phi_{j+3}$, $j=1,2,3$, the symmetry
acts by $W_j \to e^{ i \alpha} W_j $.
Planar diagrams in non-commutative theories are
particularly simple. The only difference between the planar
diagrams in the ordinary theory and the theory  with the star
product is an overall phase depending only on the external
particles \cite{filk} (see also \cite{Bigatti}). Thus, if we consider a quantity such as the
free energy, which does not involve any external particles, then
the final result is the same as in the ordinary theory.

The gravity duals of such theories are easy to construct.
We simply need to do a particular transformation on the gravity
solution. This transformation has the following origin. In the
original solution we have two commuting isometry directions, $x^-$
and an angle $\varphi$ that is conjugate to the charge $Q$ we
defined above. If we naively imagine doing a dimensional reduction
of the ten dimensional gravity theory on these two dimensions we
get a theory in eight dimensions that has an $SL(2,\R)$ symmetry.
This $SL(2,\R)$ symmetry is a symmetry in the eight dimensional
gravity theory but it is not a symmetry of the ten dimensional
configuration. In fact, it maps ordinary solutions into the
solutions we want. This transformation is the following simple set
of steps. We first do a T-duality in the direction $\varphi$ to a
T-dual direction $\tilde \varphi$. We then do a shift of
coordinates $ x^- \to x^- + \sigma \tilde \varphi$. Finally,  we do
a T-duality again on $\tilde \varphi$ into $\varphi$.
 The step where we did a shift of $x^-$ is not a symmetry of the
 theory since $\tilde \varphi$ is periodic but $x^-$ is not.
 Nevertheless this operation generates a new solution which is
 another solution of the gravity equations which is not equivalent
 via legal U-dualities to the original one. This procedure was
 applied in \cite{dipolegravity} to obtain the corresponding
 gravity solutions. We perform these steps more explicitly in
 appendix \ref{BH}. We start with AdS$_5\times S^5$, where the AdS$_5$ metric
 is written as in (\ref{adsmetr}).  The final type IIB background that we obtain through this
 procedure is
 \bea \label{metricdip}\label{sch:metric}
 \diff s^2 & = & - \sigma^2 r^4  ( \diff x^+)^2 + r^2\left[ - \diff x^+ \diff x^- + \diff \vec x\,{}^2\right] + {\diff r^2 \over r^2}  + \diff s^2 ({\cp^2}) + \eta^2, \nn\\
 B^{NS} & = & \, \sigma r^2  \diff x^+ \wedge \eta~,
\eea
and the rest of the fields have the same form as they had
before we applied the transformation.
Here $\eta = \diff \phi + P$ is the one-form present in any Sasaki--Einstein manifold, which is dual to the Reeb vector field $\de / \de \phi$. In the case of $S^5$, this generates simultaneous rotation
 of the scalars as discussed above. Indeed, in \eqref{metricdip} we can replace
the $\cp^2$ space with any  local K\"ahler-Einstein space $B_{KE}$, so that
\bea
\diff s^2 (Y)\, =\, \diff s^2({B_{KE}}) + \eta^2
\label{semetric}
\eea
is a Sasaki-Einstein metric on $Y$, where $\diff \eta /2=\omega$ is the K\"ahler two-form on $B_{KE}$.
These backgrounds do not  preserve any supersymmetry.
See Appendix \ref{schsusy}.

The five-dimensional part of this metric has the same form as the ones considered in
\cite{son,mcgreevy}. These metrics have the feature that they
break the symmetries to the Schr\"odinger symmetry even before compactifying
the direction $x^-$. However, if we do not compactify the direction $x^-$ we have
a theory with continuous eigenvalues for $P_-$ so that there is no sense in which
we have a non-relativistic system. On the other hand, if we do not compactify $x^-$
we can definitely trust the metric \nref{metricdip}, at least for values of $r$ that are not
too large.

However, we wanted to perform the DLCQ of this theory. When we compactify $x^-$ we
find that we can no longer trust computations in the metric \nref{metricdip}.
As before, we could consider a configurations with finite $P_-$. A simple case arises
when we consider a finite temperature configuration.
We can obtain the corresponding metric if we start with the black brane solution
\nref{blackthree} and perform the TsT transformation that lead to \nref{metricdip}.
The final metric in the Einstein frame is
(see appendix \ref{BH}  for details)
\bea
&& \diff s^2 \, = \,  \ex^{\tfrac{3}{2}\Phi} r^2
 \bigg[ \left( -1  + \frac{r_0^4}{2r^4}\right) \, \diff x^+ \diff x^- + \frac{r_0^4}{4r^4}
 \left( \lambda^{2} (\diff x^-)^2 +  \lambda^{-2} (\diff x^+)^2 \right)\nn\\[2mm]
&& \qquad \qquad \qquad  - \,\sigma^2 r^2 \, \left(1-{r_0^4\over r^4}\right)  (\diff x^+)^2 \bigg]
 +   \ex^{-\tfrac{\Phi}{2}} r^2 \bigg[   \frac{1}{r^4-r_0^4}\, \diff r ^2  + \diff \vec{x}^2 \bigg]\nn\\[2mm]
&& \qquad \qquad \qquad +\, \ex^{-\tfrac{\Phi}{2}}\diff s^2 (B_{KE}) + \ex^{\tfrac{3}{2}\Phi} \eta^2
\label{sch:bh}
\eea
with dilaton and $B$-field given by
\bea
\ex^{-2\Phi} & = & 1 + \sigma^2 \lambda^{2} \frac{r_0^4} {r^2}   ~,\nn \\[2mm]
B^{NS} & = & \sigma  \frac{r^2}{2}\ex^{2\Phi}
\left[ \left(2 - \frac{r_0^4}{r^4}  \right)\, \diff x^+ - r_0^4 \lambda^{2}\,  \diff x^- \right]
\wedge \eta  ~,\label{sch:fields}
\eea
where $\sigma$ is the parameter used in the shift, $x^-\to x^- +\sigma\tilde \varphi$.
By construction, the solution has the desired asymptotic form at infinity as $r \to \infty$.
However, there is now a horizon at some finite value
of the coordinate $r$, inherited from the asymptotically-AdS black three-brane metric \nref{blackthree}.

We can now compute the energy and momentum of this solution, as well as its entropy.
All these results are independent of $\sigma$, which is related to the non-commutativity
parameter. In fact we simply get the results in \nref{pplusminus}, \nref{temperature},
\nref{Entropy}.  The reason is that the TsT transformation is a symmetry of the eight
dimensional gravity theory and this eight dimensional theory is all that we are using
for computing these quantities\footnote{We are also using the fact that the TsT transformation
leaves invariant
 the eight dimensional Kaluza-Klein gauge fields associated to the two charges.}.
 See Appendix \ref{thermodetails} for a more detailed explanation.
As explained in the beginning of this subsection,
 this is due in the field theory language to the
fact that the planar diagrams are the same in the two theories. Of course, it is not a
symmetry of the full theory, so we expect that at order $1/k^2$ the two theories would
yield different answers, where $k$ is the rank of the gauge group.

\section{Consistent truncations}
\label{consistent}

In this section we construct two different
consistent truncations of type IIB supergravity
with massive vector fields. As we will see, one of them  admits the black hole background
 \nref{sch:bh}, \nref{sch:fields},  among its solutions.
  Usually consistent truncations involve massless vector fields.
These truncations can be used to generate other solutions which have the same symmetries
as certain conformal quantum mechanical theories. We think that these truncations are
interesting in their own right and could perhaps be useful for the construction of
other solutions.
We have two truncations which we will discuss in turn. Both are consistent
truncations of the type IIB  equations for the bosonic fields. It might be possible that they
can be supersymmetrized.
We will state here the final results and leave details to appendix \ref{detailsreduction}.

Before proceeding, let us recall how the backgrounds with non-relativistic conformal symmetry
\bea
\diff s^2(M_{z}) \, =\, -\sigma^2 r^{2z} (\diff x^+)^2 + \frac{ \diff r^2}{r^2}
+ r^2\left( - \diff x^+\diff x^- +\diff \vec{x}\,{}^2  \right)
\label{nrmetric}
\eea
arise from  AdS gravity with a massive gauge field.
This metric is a deformation of the AdS$_{d+3}$ metric and
clearly is only invariant under a subgroup of $SO(d+2,2)$,
which  for generic values of $z$  is  the non-relativistic dilatation
group, enhanced to the Schr\"odinger group for $z=2$. See Appendix \ref{conservationlaw}.
These metrics are solutions to the equations of motion derived from the action \cite{son,mcgreevy}
\bea
S \, =\, \int \diff^{d+2}x \, \diff r \sqrt{-g}\left(R - 2\Lambda -\frac{1}{4} F_{\mu\nu} F^{\mu\nu} -\frac{m^2}{2} A_\mu A^\mu
\right)~.
\label{firstaction}
\eea
In particular, the ansatz $A_+ \propto r^{z}$ solves the equations of motion, provided
\bea
\Lambda \, =\, -\frac{1}{2}(d+1)(d+2)~, \qquad \qquad\quad
m^2 \, =\, z(z+d)~.
\eea
In the following subsections we will describe consistent truncations of type IIB supergravity, containing
the action  \nref{firstaction}, where  $d=2$, and $z=2,4$ respectively.

\subsection{Non-linear Kaluza-Klein reduction with $m^2=8$}
\label{KKreduction}
\label{schbg}

The ansatz we consider here is suggested by the observation that in any background of the type
AdS$_5\times Y$, where
$Y$ is a Sasaki-Einstein manifold, there exist certain Kaluza-Klein vector fields with mass
squared precisely $m^2=8$.
These arise from standard Kaluza-Klein reduction
of the NS $B$ field and the RR $C_2$ field along  the one-form $\eta=\diff \phi + P$ that exists
on $Y$ \cite{Martelli:2008cm}.

We then consider type IIB supergravity where only the metric, the dilaton $\Phi$,
the five-form field $F_5$
and the NSNS three-form $H=\diff B$ are non-trivial.
The equations of motion in the Einstein frame
are\footnote{We use capital Roman letters  for the ten-dimensional indices.}
\begin{align}
R_{MN}&=\frac12\partial_M\Phi\partial_N\Phi
 +\frac1{96}F_{MABCD}F_{N}{}^{ABCD}\nn \\
&\qquad +\frac14  \ex^{-\Phi} (H_{MAB}H_{N}{}^{AB} -\frac1{12}g_{MN} H_{ABC}H^{ABC})~,\label{5d:metricEOM}\\
\square_{10}\Phi &= -\frac1{12}\ex^{-\Phi}H_{ABC}H^{ABC}~,\nn\\
\diff (\ex^{-\Phi} {\star} H) &=0~,\nn\\
F_5&=\star F_5,\qquad \diff F_5=0~, \qquad H \wedge F_5=0~,
\label{5d:miscEOM}
\end{align}
where $\square_{10}=\nabla^M\partial_M$.
The final equation arises from the equation of motion of the RR 3-form,
which has been set to zero. Our ansatz  is
\begin{align}
\diff s^2_{10} &=\ex^{-\frac23(4U+V)} \diff s^2(M)
+\ex^{2U} \diff s^2(B_\text{KE}) + \ex^{2V} \eta^2~, \nn \\
B&=A\wedge \eta + \theta \, \omega~, \nn\\
F_5&=(1+\star)G_{5}\qquad\text{where}\quad G_{5}=4\ex^{-4U-V} \vol(M)~,\label{5d:ansatz}
\end{align}
where $\diff s^2(B_{KE})+\eta^2$ is a Sasaki-Einstein metric.
Here, $\diff s^2(M)$ is the Lorentzian metric of the  five-dimensional part $M$ and  $\vol(M)$
denotes its volume form.
We use indices $x^a$ ($a=0,\ldots,4$) to denote the directions along $M$
and $x^i$ ($i=1,\ldots,4$) for those along $B_{KE}$.
$U$, $V$, $\theta$ and $\Phi$
are scalar functions, and $A$ is a one-form on $M$, respectively.
The warp factor in front of
$\diff s^2(M)$ is inserted to obtain
 the reduced equations in the five-dimensional Einstein frame.
This ansatz is closely related to the one used in \cite{hep-th/0002159},
where they had scalars $U$, $V$, $\theta$ and $\Phi$, but not the gauge field $A$.
They also had non-trivial RR 2-form, which we do not have.

Our choices for $A$ and $F_5$ satisfy  the equations  \eqref{5d:miscEOM}
automatically. Then defining  $F=\diff A$,
the field strength $H=\diff B$ of the two-form $B$ reads
  \begin{equation}
H\, =\, F\wedge \eta - (2A-\diff\theta)\wedge \omega~.
\end{equation}
The gauge transformation of the $B$-field
$B\to  B+\diff(\chi\wedge \omega)$
with a function $\chi$ on $M$ induces the five dimensional gauge transformation
\begin{equation}
A\, \to\,  A+\diff\chi~,\qquad \theta\to\theta-2\chi~,
\end{equation} which shows that
$\theta$ is the St\"uckelberg field giving mass to the gauge field $A$.
We choose the gauge $\theta=0$ for simplicity.

The equations of motion
\nref{5d:miscEOM}
follow from the Type IIB  action
\begin{multline}
S\, =\, \frac12\int \diff^{10}x \sqrt{-g} \Bigl[R
-\frac12\partial_A\Phi\partial^A\Phi\\
-\frac1{2\cdot 3!}\ex^{-\Phi}
H_{ABC}H^{ABC}-\frac1{2\cdot 5!}G_{(5)\, ABCDE}G_{(5)}{}^{ABCDE}
\Bigr]\label{5d:10daction}
\end{multline}
by varying $g_{MN}$, $B$ and $\Phi$.
Inserting the ansatz  \eqref{5d:ansatz}   in \eqref{5d:10daction} and integrating over the internal directions,
we obtain the five-dimensional action
\begin{multline}
S\, =\, \frac12\int \diff^5x\sqrt{-g} \Bigl[ R
+24 \ex^{-u-4v}-4 \ex^{-6u-4v} - 8 \ex^{-10v}
-5\partial_a u\partial^a u-\frac{15}{2}\partial_a v\partial^a v\\[2mm]
-\frac12\partial_a\Phi\partial^a \Phi
-\frac14 \ex^{-\Phi+4u+v}F_{ab}F^{ab}- 4 \ex^{-\Phi-2u-3v}A_a A^a\Bigr]~,\quad
\label{5d:5daction}
\end{multline}
where  we defined $u=\frac25(U-V)$ and $v=\frac{4}{15}(4U+V)$ to diagonalize the
kinetic terms for the scalars.
The five-dimensional equations of motion which follow from the action above are
\begin{align}
R_{ab}&=-\frac13(24 \ex^{-u-4v}-4 \ex^{-6u-4v} - 8 \ex^{-10v} )g_{ab} \nn \\
& \qquad +\frac12\partial_a\Phi\partial_b \Phi +
5\partial_a u\partial_b u+\frac{15}{2}\partial_a v\partial_b v\nn \\
&\qquad +\frac12\ex^{-\Phi+4u+v}(F_{ac}F_{b}^{c}-\frac16 g_{ab}F_{cd}F^{cd})
+ 4 g_{ab}\ex^{-\Phi-2u-3v}A_c A^c,\nn\\
\diff (\ex^{-\Phi+4u+v} {\star_5 F}) &=-8 \ex^{-\Phi-2u-3v} \star_5 A,\nn\\
\Box_5 \Phi&=-\frac14\ex^{-\Phi+4u+v}F_{ab}F^{ab}-4\ex^{-\Phi-2u-3v}A_a A^a,\nn\\
10 \Box_5  u&=24(\ex^{-u-4v}-\ex^{-6u-4v})\nn\\
&\qquad +\ex^{-\Phi+4u+v}F_{ab}F^{ab}-8\ex^{-\Phi-2u-3v}A_a A^a,\nn\\
15  \Box_5   v&=16(6 \ex^{-u-4v}-\ex^{-6u-4v}-5 \ex^{-10v})\nn\\
&\qquad +\frac14\ex^{-\Phi+4u+v}F_{ab}F^{ab}-12\ex^{-\Phi-2u-3v}A_a A^a \label{5d:EOMs}
\end{align}
where $\square_5=\nabla^a \partial_a$ is the five-dimensional d'Alembertian.

Interestingly, this non-linear Kaluza-Klein reduction is consistent: any solution
to the five-dimensional equations of motion \nref{5d:EOMs} can be lifted to a solution of
the ten-dimensional equations of motion of type IIB supergravity,
 using the ansatz \eqref{5d:ansatz}.
The details can be found in Appendix~\ref{KKdetails}.
The masses of the excitations around the AdS$_5$
background are
 \begin{equation}
m^2_A\,=\, 8~,\qquad m^2_\Phi\,=\,0~,\qquad m^2_u\,=\, 12~,\quad \text{and}\quad
m^2_v\,=\, 32~.
\end{equation}
 Notice that one can write a ``superpotential'' (in the Hamilton-Jacobi sense)
for the scalar  fields $u,v$ \cite{hep-th/0002159}.
This may be useful for studying holographic renormalisation of the system, see e.g.
 \cite{de Haro:2000xn,Campos:2000yu,Martelli:2002sp,Papadimitriou:2004ap}.


\subsubsection*{Background with $z=2$}

As promised, we can now see that a
particular solution to the equations of motion
\eqref{5d:EOMs}
is  obtained by setting $\Phi=u=v=0$,
taking (\ref{nrmetric}) with $z=2$ as metric on $M$ and the gauge field to be
\bea
A \, = \, \sigma r^2 \diff x^+~.
\eea
Indeed, setting to zero the scalar fields,
 the action  \eqref{5d:5daction} reduces to the action
 \eqref{firstaction} discussed in \cite{son,mcgreevy}.
 We then recover in a different way the solution
 \nref{metricdip} previously obtained via the TsT transformation.
As already mentioned, this solution does not preserve any supersymmetry.
The derivation of this is relegated to Appendix~\ref{schsusy}.

In fact, it can be checked that also the black hole background \nref{sch:bh},  \nref{sch:fields} arises as a solution to the equations of motion \eqref{5d:EOMs}, with non trivial dilaton and scalar fields $u,v$. This provides an explicit  check that   this is indeed a
solution of type IIB supergravity.
There may be other interesting solutions to the equations \eqref{5d:EOMs}, with non-relativistic symmetry or otherwise. We leave a more complete analysis for future work.

\subsection{Non-linear Kaluza-Klein reduction with $m^2=24$}

\label{dilbg}

Following the logic of Sec. \ref{KKreduction}, it is natural to  consider  other
special Kaluza-Klein modes on $Y$, that could be promoted to full non-linear consistent reductions.
After having considered two-form modes, we should  look at Kaluza-Klein
modes coming from the RR four-form potential,
as well as the metric. In fact, it is well known that these modes mix and come in pairs \cite{kim}
with  mass eigenvalues
\bea
m^2_\pm \, = \, \mu + 4 \pm 4 \sqrt{\mu+1} ~,
\eea
where $\mu$ is the eigenvalue of the Laplacian
on the massive three-form $\omega_3$ in $Y$ on which the
RR potential is reduced, that is $C_4 = A \wedge \omega_3$.
For any Killing vector field in $Y$, there exists a mode with eigenvalue
 $\mu=8$  \cite{Martelli:2008cm,Benvenuti:2006xg}.
Thus, for each Killing vector, one obtains a massless mode as expected, but also a massive mode,
with  mass $m^2_+=24$.
It has been shown \cite{BuchelLiu,Gauntlett:2007ma} that the massless KK mode associated to the Reeb Killing vector
may be promoted to a non-linear truncation, yielding minimal gauged supergravity. In the following, we will
demonstrate that in fact it is possible to include in the truncation also the massive gauge field, at least at the level of bosonic fields.

We then consider a background with only the metric and the five-form field non-trivial, with
the following ansatz:
\begin{align}
\diff s^2&=\ex^{-\frac23(4U+V)}\diff s^2(M)+\ex^{2U}\diff s^2(B_\mathrm{KE})+\ex^{2V}(\eta+
\cA)^2~,\\
F_5&= (1+\star_{10})
\left[2\omega^2\wedge (\eta+\cA)+2\omega^2 \wedge \bA-\omega\wedge(\eta+\cA)
\wedge \bbF\right]~.
\end{align}
Here we gauged the Reeb isometry direction $\eta$ by a connection one-form
$\cA$ on $M$, and we will denote its curvature by $\cF=\diff \cA$.
The ansatz for $F_5$, which has one-form $\bA$ and two-form $\bbF$ on $M$,
requires some explanation.
The $2\omega^2\wedge \eta$ term is chosen to have $\int_Y F_5=4$
which is the value for the AdS$_5\times Y$ background.
The structure $2\omega^2\wedge \bA-\omega\wedge \eta \wedge \bbF$
is suggested by the discussion above that we should take $C_4\sim \omega\wedge\eta\wedge \bA$,
therefore the combination
\begin{equation}
\diff(\omega\wedge\eta\wedge \bA) \,=\,
2\omega^2\wedge\bA-\omega\wedge\eta\wedge \bF
\end{equation}
with $\bF=\diff \bA$ should be in the expansion of $F_5$.
The ansatz is then made gauge-invariant by the
replacement $\eta\to\eta+\cA$, which imposes
\begin{equation}
\bbF\,=\, \cF+\bF
\end{equation}
from the $\omega^2$ component of the closure of $F_5$.
This ansatz gives again a consistent reduction,
which is an extension of the consistent reduction to the (bosonic sector of the) minimal
gauged supergravity \cite{BuchelLiu,Gauntlett:2007ma}.
The details of the calculation are explained in Appendix \ref{msq24detail}.

Here we only discuss the equations of motion of the vector fields $\cA$ and $\bA$,
one of which comes from the $\omega$ component of $\diff F_5=0$,
\begin{align}
\diff ( \ex^{-\frac43(U+V)}\star_5 \bbF ) &=-8\ex^{-8U}\star_5 \bA +\cF\wedge\bbF~.
\end{align}
In addition,  the 10d Einstein equation in \nref{5d:miscEOM},
with $(A,B)=(a, \phi)$ gives the relation
\begin{equation}
\diff (\ex^{\frac83(U+V)}\star_5 \cF )\, =\, 16\ex^{-8U}\star_5 \bA+\bbF\wedge \bbF~.
\end{equation}
These two equations, as well as the equations for the 5d metric and the scalars,
 follow from the action
\begin{multline}
S=\frac12\int \diff^5x\sqrt{-g} \Bigl[ R
+24 \ex^{-u-4v}-4 \ex^{-6u-4v} - 8 \ex^{-10v}
-5\partial_a u\partial^a u-\frac{15}{2}\partial_a v\partial^a v\\[2mm]
-\frac14 \ex^{-4u+4v}\cF_{ab}\cF^{ab}
- \frac12  \ex^{2u-2v} \bbF_{ab} \bbF^{ab} -8\ex^{-4u-6v} \bA_a\bA^a \Bigr]
+\frac12\int \cA\wedge \bbF \wedge \bbF~,
\label{5daction:m2=24}
\end{multline}
where $u=\frac25(U-V)$ and $v=\frac{4}{15}(4U+V)$.
As we explain in Appendix~\ref{msq24detail}, this is a consistent reduction.

When $u=v=0$, the kinetic term of the gauge fields can be diagonalized, with the full action
becoming
\begin{multline}
S_\text{vector}=\frac12\int \diff^5 x \sqrt{-g}\left[
 -\frac34(\cF+\frac23\bF)_{ab}(\cF+\frac23\bF)^{ab}
 -\frac1{6}\bF_{ab}\bF^{ab}-8 \bA_a\bA^a
 \right] + S_{CS}
\end{multline}
where
\bea
 S_{CS} \, =\, \frac12\int \, \cA\wedge(\cF+\bF)\wedge(\cF+\bF)~.
\eea
Therefore we have one massless mode $\cA+\frac23\bA$,
which appears in the ordinary gauged supergravity,
and one massive mode $\bA$ with $m_{\bA}^2=24$.

\subsubsection*{Background with $z=4$}

As an example, let us construct a simple solution to the equations of motion which follow from \eqref{5daction:m2=24}.
We can set $u=v=0$ consistently if $|\bA|^2=|\cF|^2=|\bbF|^2=0$,
then we can choose $\bA=-\frac32 \cA$ to set the massless gauge field to be zero.
The action then becomes \eqref{firstaction} with the null condition $|A|^2=|F|^2=0$ as was
the case for the  background with $z=2$.
Therefore one solution is
obtained  by taking (\ref{nrmetric}) with $z=4$ as metric on $M$ and
\bea
\cA \, = \, \sigma\, r^4\diff x^+~.
\eea
We then find that the following is a solutions of type IIB supergravity
\bea
&&\diff s^2 \, = \,  - \sigma^2  r^8(\diff x^+)^2 + \frac{\diff r^2}{r^2}
 + r^2\left(-\diff x^+\diff x^- +\diff \vec{x}\,{}^2 \right) \nn\\[1.5mm]
&& \qquad \qquad \qquad  +
 \diff s^2 (B_\mathrm{KE}) + (\eta + \sigma r^4 \diff x^+)^2~,\label{harry}\\[1.5mm]
&& F_5 \, = \, 4 \vol (M) + 2 \omega \wedge \omega \wedge \eta
\nn\\
&& \qquad \qquad \qquad -  \sigma  \diff x^+ \wedge
\left[ \omega_C \wedge \omega_C + \diff x^1 \wedge \diff x^2 \wedge \diff (r^6\eta)
\right]~,
\label{dilmetric}
\eea
where recall $\diff \eta \, = \, 2 \omega$ and we defined
\bea
\omega_{C} \, = \, \frac{1}{2} \diff (r^2 \eta)~,
\label{omegac}
\eea
which is  the K\"ahler two-form on the Calabi-Yau cone $C(Y)$,  $\diff s^2(C(Y)) = \diff r^2+r^2 \diff s^2(Y)$.
Notice the analogy with the background \nref{metricdip}. In particular, we can think of the background here as a deformation
of AdS$_5\times Y$, with deformation parameter $\sigma$. However, in this case, the solution cannot
be related to an AdS background by a simple TsT transformation.

This is a background enjoying  non-relativistic symmetry with exponent $z=4$.
Based on the analogy with the $z=2$ case, it is plausible that black hole geometries  with \nref{harry},
\nref{dilmetric} as asymptotic boundary conditions arise as solutions to the equations with non-zero
scalar fields. Since these are not accessible by simple duality transformations, it would be interesting
to try to find such black hole solutions, which should be dual to non-relativistic systems with $z=4$.
It would also be nice to understand the field theory duals of these solutions.
Moreover, it is possible that many more solutions could be found, with completely different applications.


\section{Conclusions}\label{conclusions}

In this paper we have considered some conformal quantum mechanical systems that arise by
performing the DLCQ of conformal field theories. We have considered several aspects of the
gravitational geometries that arise via this procedure. We have emphasized that we cannot
trust the gravity solutions that we get by performing the DLCQ of AdS space. The correct
DLCQ description is usually more involved. Fortunately, in cases with sufficiently high
light cone momentum $N$, or high momentum density, we can trust the geometry at least in some
region. Situations were the metric can only be trusted
in some region of the geometry are common. For example, in the
gravity dual of D0 brane quantum mechanics the geometry can be trusted only in certain
regions,  see \cite{Itzhaki}\footnote{In this case the gravity results were compared with numerical simulations in
\cite{Anagnostopoulos:2007fw,TW}.}. In particular, this allows to compute certain correlation functions,
as long as the length scales at which we compute them lie in the region where the computation
is dominated by the gravity result.
On the other hand, if $N$ is small we cannot trust results computed on the AdS metric with
$x^-$ identified. Of course, if the modes we consider are BPS, it is possible that we get the
right answer even if we do the naive computation.

We considered thermal configurations and we concluded that we can trust the geometries
near the horizon if the momentum density is large enough. In this way we computed the thermodynamic
properties of the dual quantum mechanical theories.
Thus, we conclude that for the solutions with translationally
invariant horizons, in which the hydrodynamic regime makes sense,
the transport properties will have the usual gravity values. In particular, it is simple
to see that the bulk
viscosity coefficient vanishes, as dictated by conformal invariance \cite{nishidason,bulk},
and  the shear viscosity has its universal value \cite{pss,pss2}.

  We have given one concrete example of
 a quantum mechanical theory that we can analyze using these gravity duals. This is the conformal
 quantum mechanical theory describing the DLCQ of M5 branes \cite{ABS}. This is a sigma model
 whose target space is the moduli space of $N$ instantons in $U(k)$ and it is given in terms
 of the ADHM construction reviewed in Sec. \ref{M5}. Naively one might have expected
 that the quantum mechanical description would involve a system of $N$ particles moving
 in four spatial dimensions. However, the
 quantum mechanical theory has $4 Nk$ variables, but realizing the Schr\"odinger group in
 4+1 dimensions \cite{ABS}.  This highlights the subtle nature of DLCQ.
   The thermal properties of this theory
 were computed in Sec. \ref{ADSseven}. For large enough $k$ and $N$ we can trust the
 gravity computations. The black hole geometries were  obtained by taking a simple limit
 of the  Kerr-AdS black holes \cite{Hawking:1998kw,AdSKerrFirstLaw}.
 The quantum mechanics theory does not look particularly similar to the fermions at unitarity,
 which was the initial motivation for \cite{son,mcgreevy}. However, it is interesting that
 we get some concrete conformal quantum mechanical theories that can be studied in this fashion.
 Maybe this point of view might shed some light on the mysterious connection between AdS$_2$ and
 conformal quantum mechanics.

 We have also observed that   some of the
  backgrounds introduced in \cite{son,mcgreevy} can be realized
 in string theory by considering the gravity duals of the dipole non-commutative deformation
 of conformal field theories \cite{dipolelightlike}. In this case the gravity description with
 and without the deformation are very closely related. In fact, if we consider observables with zero
 external charge and momentum the results are the same as in the ordinary theory.

 We have also found certain consistent Kaluza-Klein reductions of type IIB supergravity involving
 massive vector fields. These reductions have also allowed us to find  an embedding for a solution
 with non relativistic conformal symmetry with a dynamical exponent $z=4$, which is different from the
 Schr\"odringer value, $z=2$. It would be nice to find the field theory interpretation of these solutions.
  These Kaluza-Klein reductions are probably useful in their own right,
 independently of any applications to the backgrounds in \cite{son,mcgreevy}.

{\bf Note added in v5} 

The dependence of the thermodynamic quantities on the temperature and chemical potential 
can be understood on the basis of scale invariance together with the fact that the theory 
has a gravity dual which is boost symmetric before we perform the light-like compactification. 
For details see appendix F. 

\section*{Acknowledgments}

The authors  benefited from discussions with  Vijay Balasubramanian
and Oren Bergman. D. M. would like
to thank the Aspen Center for Physics, for enjoyable hospitality during the completion of this work.
This work was supported by  NSF grant PHY-0503584 and DOE Grant DE-FG02-90ER40542.
The work of Y. T. is in part supported by the Carl and Toby Feinberg fellowship
at the Institute for Advanced Study.


\appendix

\section{Non-relativistic conformal symmetries}
\label{conservationlaw}

Let us summarise here the non-relativistic conformal groups and their embeddings
to the relativistic counterpart \cite{son,mcgreevy,ABS}.
Recall the Galilean algebra in $d$ spatial dimensions has
the Hamiltonian $H$, momenta $P_i$, rotations $M_{ij}$,
Galilean boosts $K_i$ and the mass operator $M$ which commutes with everything.
Less obvious commutators are
\begin{equation}
\left[ P_{i}, K_j \right] \, = \, - i \delta_{ij} M ~,\qquad
\left[ H,K_i \right] \, =\, - i P_i ~.
\end{equation}
This algebra can be embedded in the Poincar\'e algebra with $d+1$ spatial dimensions
and one timelike dimension. We denote its generator by putting a tilde on the symbols,
thus $\tilde M^{\mu\nu}$ are  the rotations and $\tilde P^\mu$ are the translations.
The Greek indices take values $0,1,\dots d+1$.
The Galilean algebra is obtained by  introducing the light-cone coordinates
\bea
x^\pm \, =\, x^0\pm x^{d+1}
\eea and retaining the subalgebra commuting with  $\tilde P_-$, which is
interpreted as the mass operator $M$.
The identification is given by
\be
M\,=\, -\tilde P_-~, \quad
H \, = \, -\tilde P_+~, \quad P_i \,= \, \tilde P_i~, \quad M_{ij} \, =\, \tilde M_{ij}~, \quad
K_i \, =\, \tilde M_{-i}~.\label{identification}
\ee
This embedding is well-known in the context of discrete light-cone quantisation,
as recalled in the main text.

The Galilean algebra may be extended to the non-relativistic conformal group,
by including a dilatation generator $D$,  whose non-zero commutators are
\begin{align}
\left[ D, P_i \right] \, &= \, -iP_i~, &
\left[ D, H \right] \, &= \, -iz H~, \nn\\
\left[ D, K_i \right] \, &= \, i(z-1) K_i~, &
\left[ D, M \right] \, &= \, i(z-2)M~.
\end{align}
The constant $z$ is referred to as the ``dynamical exponent'' in the condensed matter literature, and reflects the freedom
to scale differently time and space coordinates \cite{mcgreevy}. In the special case that $z=2$, there exists a further extension of the group, obtained by adding a
special conformal transformation with generator $C$. The new non-zero commutators are then given by
\bea
\left[ C, P_i \right] \, = \, i K_i~, \qquad
\left[ D, C \right] \, = \, 2i C~, \qquad
\left[ H , C\right] \, = \, i D~,
\eea
and the resulting group is called  the  Schr\"odinger group.  Notice that this contains an $SL(2,\R)$ subgroup generated
by $H,D,C$. This can be conveniently presented in terms of an ``oscillator Hamiltonian'' $H_{osc}=L_0=\frac{1}{2}(H+C)$ and the
raising/lowering operators $L_{\pm}= \frac{1}{2}(H-C \mp i D)$ \cite{DFF}.

The Schr\"odinger algebra is obtained, just as in the case of the Galilean algebra,
as the  sub-algebra of generators commuting with $\tilde P_-$.
One identifies the generators  as follows:
\bea
D \, = \, \tilde D + B~, \qquad
C \, =\, -\tilde K_-    ~.
\eea where $\tilde D$ and $\tilde K^\mu$ are the dilatation and special conformal transformations of the relativistic conformal group. $B=2\tilde M_{-+}$ is the generator of the boost normalised so that
$x^\pm$  has eigenvalue $\pm1$.
For $z\ne 2$, the non-relativistic dilatation is realized as the combination
\begin{equation}
D\, =\, \tilde D+(z-1)B
\end{equation} in the relativistic conformal algebra.


\section{Black holes with plane wave asymptotics}
\label{BHplane}

\subsection{Harmonic potential metric}

We start with the one-parameter  five dimensional Kerr-AdS metric
originally found in \cite{Hawking:1998kw}, and write this in the form presented in \cite{AdSKerrFirstLaw}. Namely, the metric is
\bea
&&\diff s^2\,=\, -\frac{\Delta_r}{r^2+a^2\sin^2\theta}(\diff t-\frac{a}{\Xi}\cos^2\theta \diff \phi)^2
+\frac{r^2+a^2\sin^2\theta}{\Delta_r}\diff r^2+\label{AdSKerr}\\[2mm]
&&\qquad \quad  +\frac{r^2+a^2\sin^2\theta}{\Delta_\theta}\diff \theta^2
 \quad +\frac{\Delta_\theta\cos^2\theta}{r^2+a^2\sin^2\theta}(a\diff t-\frac{r^2+a^2}{\Xi}\diff\phi)^2
+r^2\sin^2\theta \diff\psi^2~.\nn
\eea
where
\begin{equation}
\Delta_r=(r^2+a^2)(1+r^2)-2m,\qquad
\Delta_\theta=1-a^2\sin^2\theta,\qquad
\Xi=1-a^2.
\end{equation}
Note that our $\theta$ is their $\pi/2-\theta$, \emph{i.e.} $\sin\theta$ and $\cos\theta$ are flipped
with respect to theirs. This metric has $J_\phi\ne 0$, $J_\psi=0$.
We need to have $|a|<1$ to have  space-like $\theta$ direction.
The energy and the angular momenta are \cite{AdSKerrFirstLaw} (with $R^3_{AdS}/G_N^5=1$)
\begin{equation}
E\,=\,\frac{\pi m (3-a^2)}{4(1-a^2)^2}~,\qquad
J_\phi\,=\,\frac{\pi m a}{4(1-a^2)}~,\qquad
J_\psi \,=\,0~.
\label{AdSKerrEnergy}
\end{equation}

This metric is asymptotically  AdS$_5$, but in the coordinate system above the boundary is a rotating
Einstein universe. To display the  $\R\times S^3$ boundary, we need the coordinate change
\cite{Hawking:1998kw}
\begin{equation}
(1-a^2) \hat r^2 \cos^2\hat\theta \,=\, (r^2+a^2)\cos^2\theta,\qquad
\hat r^2\sin^2\hat\theta\,=\,r^2 \sin^2\theta~,\qquad
\hat\phi\,=\,\phi+a t~.
\label{asymptoticchange}
\end{equation}
Then the asymptotic form becomes
\bea
&&\diff s^2\,=\, -(1+{\hat r }^2)\diff t^2+\frac{\diff {\hat r }^2}{1+{\hat r }^2-
\frac{2m}{\Delta_{\hat\theta}}} +
{\hat r }^2\big(\diff\hat\theta^2+\cos^2\hat\theta\diff\hat\phi^2+\sin^2\hat\theta\diff\hat\psi^2\big)\nn\\
&&\qquad \quad +\frac{2m}{{\hat r }^2(1-a^2\sin^2\hat\theta) ^3}
(\diff t-a\cos^2\hat\theta \diff\hat\phi)^2+\cdots.
\eea
The boundary stress-energy tensor in this coordinate system is calculated in \cite{Bhattacharyya:2007vs}.

Since we  want to obtain the asymptotic plane wave  metric from the $\R \times S^3$ metric,  we perform
the limit \eqref{RxS3planewave}.  Namely, we define
 \be \label{coordch}
  t \,=\, x^+ ~,~~~~~
  t-  \hat \phi \, =\,{  x^- \over 2 R^2} , ~~~
  \hat \theta \,= \,\  {\rho \over R} ~,~~~~ \hat r \,=\, R   y~.
 \ee
and  let $R \to \infty$, keeping
$x^\pm$ , $\rho$ and $  y$ fixed.
We also need to scale $E$ and $J$ accordingly. The energy is conjugate to $t$ translations and
the angular momentum $J$ to $\hat \phi$ translations. The coordinate change \nref{coordch}
implies that $- P_+ = E-J$ and $-P_- = { J \over 2 R^2 }$. However, in addition we need
to take into account that when we compactify $x^-$ we are effectively shrinking the $\hat \phi$
circle from radius one to radius $r^-/(2 R^2)$. This rescales the energy and the angular momentum
since the configuration is translation invariant along $\hat \phi$. The final expressions are
\be \label{ppmgen}
-P_+ = { r^- \over 2 R^2 } ( E- J ) ~,~~~~~~~~~ - P_- = { r^- \over 2 R^2 } { J \over 2 R^2 }.
\ee
We see from \nref{AdSKerrEnergy} that  in order to get a finite limit we need to take $a\to 1$.
Thus we end up doing the scalings
 \be \label{finalscal}
  t \,=\, x^+ ~,~~~~~~ \phi\, =\, - { x^- \over 2 R^2 } + (1-a) x^+  ~,~~~~~~
  1-a \,=\, {1 \over 2 \lambda R^2} ~,~~~~~R \to \infty
 \ee
  and $r$ and $\theta $ are not scaled. Asymptotically, they are
  related to the scaled variables $\rho$ and $ y$ by
  \be \label{changc}
\lambda^{-1}  y^2\, =\, (r^2 +1) \cos^2 \theta ~,~~~~~~~   y^2  \rho^2 \,=\, r^2 \sin^2 \theta.
  \ee
  The limiting metric was given in  \nref{bhpw}.
We can see that the $\lambda$ dependence can be removed by rescaling $x^-$,
but it is convenient to keep $\lambda$ for some purposes.
We can compute $P_\pm$ by taking the limits of   \nref{ppmgen}. We find
\be
- P_- \,=\, { \pi \lambda^2 r^-} { m \over 8 } ~,~~~~~~ -P_+ \,=\, \pi \lambda r^- { m \over 4 }~,
\ee
which lead to \nref{partdens}.

We have not found a simple  analytic form of the metric above in the asymptotically plane
wave coordinates, but for large $y$ we can expand it as in  \nref{ppBH}.
The boundary stress-energy tensor is then given by
$T_{ab}=\frac {\lambda m}{(1+\lambda \rho^2)^3} S_{ab}$ where
\bea
&& S_{++}\,=\,1+3\lambda\rho^2+3\lambda^2\rho^4~,\quad
S_{+-}\,=\,\frac{\lambda(1+3\lambda\rho^2)}2~,\quad
S_{--}\,=\,\lambda^2~,\nn\\
&& S_{\rho\rho}\,=\,\lambda(1+\lambda\rho^2~),\qquad\qquad
S_{\psi\psi}\,=\,\rho^2S_{\rho\rho}~,
\eea
which is traceless, as it should be.

In order to obtain the metric for the case of the black hole in AdS$_7$ we simply
need to start from the general rotating black holes in seven dimension as written in
 \cite{AdSKerrFirstLaw} (see
also \cite{kerrds}). The metrics are very similar up to the replacement $m \to m/r^2$ and
$\diff\psi^2 \to \diff \Omega_3^2$. The scalings we need to do are precisely as in \nref{finalscal}
and the resulting metric is \nref{bhpwseven}.
 The temperature has a different expression because we replaced the constant parameter $m$
 by $m/r^2$ which depends on $r$. The expressions for the temperature and chemical potential
 can be found in \cite{AdSKerrFirstLaw} and we get \nref{tempmusev} after taking the limit.
 Similarly, the expressions for the energy and angular momentum that are given in
 \cite{AdSKerrFirstLaw} (with $R_{AdS}^5/G_N^7 =1$)
 \be
 E \,=\, { m \pi^2 \over 4 (1 -a^2) } \left( { 1 \over (1 -a^2) } + { 3 \over 2 } \right)
  ~,~~~~~~~~~~~J \, =\, { m a \pi^2 \over 4 (1 -a^2)^2 }~,
  \ee
 and lead to the following values for $P_+$ and $P_-$
 \be
 -P_-\, =\, { \pi^2 \over 16 } \lambda^2 m r^- ~,~~~~~~~~-P_+ \,= \,{ \pi^2 \over 4 } \lambda m r^-~.
 \ee


\subsection{Magnetic field metric}
\label{magnetic}

We consider here a different limit, obtained starting with the general two-parameter Kerr-AdS metric
\cite{Hawking:1998kw}. In the limit, the two parameters become trivial, so for simplicity
we discuss the case that the two parameters are equal.
We start with the metric in  the form presented in  \cite{Chong:2005hr}, and set
$a=b$ and  $q=0$. We also set $g=1$ without loss of generality.
The metric is
 \begin{multline}
\diff s^2\, =\,  -\frac{1+r^2}{1-a^2}\diff t^2
+\frac{2m}{(r^2+a^2)(1-a^2)^2}\bigg[\diff t -a(\cos^2\theta \diff\phi+\sin^2\theta\diff\psi)\bigg]^2 \\[2.2mm]
+  (r^2+a^2)  \left[      \frac{\diff r^2}{\Delta_r} + \frac{1}{1-a^2} \left(
 \diff \theta^2 + \cos^2\theta \diff\phi^2+\sin^2\theta \diff\psi^2 \right)\right]~,\qquad
\end{multline}
where
\begin{align}
\Delta_r&=\frac{(r^2+a^2)^2(1+r^2)}{r^2}-2m~.
\end{align}
Note we switched  the variables $\psi $ and $\phi$  in \cite{Chong:2005hr}, which is equivalent to
the replacement $\theta\to\pi/2-\theta$.
Now let us introduce the following definitions
\begin{equation}
\theta\, =\, \frac{\rho}{R}~,\quad
t\,=\, x^+,\quad
\phi\,=\, x^+-\frac{x^-}{2R^2}~,\quad a\, =\, 1-\frac{1}{2\lambda ^2R^2}~,
\end{equation}
and then take the limit $R\to\infty$.
We then get the following result for the metric in this limit
 \begin{multline}
\diff s^2 \, =\, (1+r^2) \frac{\diff r^2}{\Delta_r}
 +\lambda^2(1+r^2)\left[- \diff x^+\diff x^- -\rho^2 (\diff x^+)^2
+ \diff\rho^2
+\rho^2\diff\psi^2\right] -(\diff x^+)^2\\[2mm]
 +\frac{2m \lambda^4}{(1+r^2)}\bigg[\frac{\diff x^-}2+ \frac{1}{2\lambda^2} \diff x^+ +\rho^2(\diff x^+-\diff \psi) \bigg]^2~,
 \qquad \qquad \qquad \quad \qquad
\end{multline}
where
\bea
\Delta_r \, =\, \frac{(1+r^2)^3}{r^2} - 2m~.
\eea
Changing the radial coordinate and the time by
\bea
y^2 \, = \, \lambda ^2 (r^2 + 1),\qquad
\hat\psi \, =\, \psi-x^+~,
\eea
we arrive at the metric
\begin{multline}
\diff s^2 \, =\,  \left(1-  \frac{2m\lambda^4}{y^4} + \frac{2m\lambda^6}{y^6}\right)^{-1}
\frac{\diff y^2}{y^2}+
y^2 \bigg[ - \diff x^+(\diff x^-  - 2\rho^2 \diff \hat\psi ) + (\diff\rho^2 +\rho^2\diff\hat\psi^2)
 \bigg] \\[2mm]
-(\diff x^+)^2 +\frac{m\lambda^6}{ 2y^2 } \left[\diff x^-+ \frac{\diff x^+}{\lambda^2} - 2\rho^2\diff \hat\psi  \right]^2~,
\label{newlimit}
\end{multline}
which is the metric \nref{maglimit} in the main text.
In the large $y \to \infty$ limit, the metric approaches the pp-wave form \eqref{adspp}.
We find that the horizon is at $y=y_H$ where
\bea
1 -  \frac{2m\lambda^4}{y_H^4} + \frac{2m\lambda^4}{y_H^6} \, =\, 0~.
\eea
which is the same as for the original  Kerr-AdS black hole metric \cite{Chong:2005hr},
 before the limit.

The boundary stress-energy tensor computed in the usual way
is then given by $T_{ab}= m S_{ab}$, where
\bea
&& S_{++}\,=\,\lambda^2 ~,\qquad
S_{+-}\,=\,\lambda^4/2 ~,\quad
S_{--}\,=\, 1 ~,\quad
S_{+ \hat\psi } \, = \, -\rho^2\lambda^4  ~, \nn\\[2mm]
&&   S_{-\hat\psi} \, =\, -2\rho^2\lambda^6~, \quad
 S_{\rho\rho}\,=\, \lambda^4,\quad
S_{\hat\psi\hat \psi}\,=\,\rho^2\lambda^4 (1+4\rho^2\lambda^2)~,
\eea
which is traceless.
It is straightforward to repeat this procedure for the seven dimensional case.

\section{Black holes with Schr\"odinger asymptotics}

\label{BH}

The background (\ref{sch:metric}) should be regarded as the gravity dual
of the  vacuum of the non-commutative dipole conformal field theory.
According to the AdS/CFT correspondence then,
several physical properties may be computed holographically from gravity duals at
finite temperature or with finite
density of states. In this appendix we address  a couple of
 issues. First   we describe how to construct a large class of explicit solutions with
asymptotic Schr\"odinger symmetry. Then
we explain how to extract physical quantities from such kind of solutions, with non-standard (e.g. non-AdS)
asymptotics. In particular, we discuss the issues arising in trying to define conserved charges (e.g. the analogue of ADM mass) and we propose how to circumvent them.
The key point is a certain duality transformation, to which we now turn.

\subsection{TsT transformation}

It turns out that the  background \eqref{sch:metric}
may be obtained from a TsT \cite{LLM} transformation
of AdS$_5\times Y$. This was discussed in  \cite{dipolelightlike} for the case of $Y=S^5$
and it extends straightforwardly to any $Y$.

Quite generally,  the TsT transformation is a solution generating technique in the context of
type II supergravity. Consider a background with two isometries
along coordinates $\varphi^1$ and $\varphi^2$. The TsT transformation consists of
three steps: firstly, we perform a T-duality along $\varphi^1$ and introduce
the dualised direction $\tilde\varphi^1$.  Secondly,
we make the coordinate shift $\varphi^2\to\varphi^2+c\tilde\varphi^1$,
and thirdly we perform another T-duality along $\tilde\varphi^1$.
The first and the third steps are dualities and
as such do not change the physics; the second step might change the periodicities
of $\tilde\varphi^1$ and $\varphi^2$, but it is guaranteed to give a new solution
at the level of supergravity equations of motion.

Let us apply this transformation to a general  type IIB solution of the form
\begin{align}
\diff s^2&=g_{ab} \diff x^a \diff x^b + \diff s^2(\tilde B) + h^2(\diff\varphi+\tilde P)^2~,\nn\\[2mm]
F_5 &= 4\left(\vol(M_5) + \vol(\tilde B)\wedge h(\diff\varphi+\tilde P)\right)~,
\end{align}
 where $\varphi$ is a Killing direction in the five-dimensional space $Y$,
$\diff s^2(\tilde B)$ is the metric of the four-dimensional space transverse to $\partial_\varphi$,
$h^2$ is the norm of $\partial_\varphi$, and $\tilde P$ is a connection one-form on $\tilde B$.
We assume all other fields are zero.

For our purposes, we T-dualise $\varphi$ into $\tilde \varphi$, then shift a light-cone coordinate
$x^-\to x^-+\sigma\tilde\varphi$, and T-dualise $\tilde\varphi$ back to $\varphi$.
The resulting solution
in the string  frame is
\begin{align}
\diff s^2&=
\left[g_{ab} \diff x^a \diff x^b
- \sigma ^2 \ex^{2(\Phi-\Phi_0)}h^2 g_{a-}g_{b-}\diff x^a \diff x^b
+ \right. \nn
\\
& \left.  + \diff s^2(\tilde B) + \ex^{2(\Phi - \Phi_0)} h^2(\diff \varphi+\tilde P )^2\right]~, \label{TsT:metric}\\
F_5&=4\left(\vol(M_5)+\vol(\tilde B)\wedge h (\diff\varphi+\tilde P)\right)~,\nn\\
B&= \sigma \ex^{2(\Phi-\Phi_0)}  h^2 g_{a-}\diff x^a\wedge (\diff \varphi+\tilde P)~,\nn\\
\ex^{-2(\Phi-\Phi_0)}& =1+\sigma ^2 h^2 g_{--}~ .\label{TsT:dilaton}
\end{align}
where $\Phi_0$ is the value of the dilaton before the duality.

If we start from the  AdS$_5\times Y$ background, where AdS$_5$ is written in the Poincar\'e patch,
 and choose the isometry $\varphi$ used to be the Reeb direction $\phi$  (or direction associated
 to the $U(1)_R$ symmetry of the ${\cal N}=1$ dual CFT), where $h^2=1$,
the metric after the TsT transformation is simply the solution
\eqref{sch:metric}.  The term $-r^4 (\diff x^+)^2$,
which is  crucial for breaking the relativistic conformal group
to the Schr\"odinger group, is generated through the term $g_{a-}g_{b-}\diff x^a \diff x^b$.
In general, the solution obtained above when $\varphi$ is the  Reeb direction
 falls inside our ansatz \eqref{5d:ansatz}
and as such gives a solution of the equations of motion  derived from the 5d action.
Thus, we have that any solution of the 5d equations, with vanishing scalars and
gauge field, may be transformed to another solution
\begin{align}
\diff s^2 (M)&=\ex^{-\tfrac{2\Phi}{3}}g_{ab} \diff x^a \diff x^b
- \ex^{\tfrac{4}{3}\Phi}\sigma^2  g_{a-}g_{b-}\diff x^a \diff x^b~,\nn\\
A&= \sigma \ex^{2\Phi}  g_{a-}\diff x^a~,\nn\\
\ex^{-2\Phi}&=1+\sigma^2  g_{--}~,\nn\\
U &=-\tfrac{1}{4}\Phi~,\qquad  V=\tfrac{3}{4}\Phi~,
\end{align}
as long as $g_{--}$ and $g_{a-}$ do not depend on the coordinate $x^-$ (and we have
set $\Phi_0=0$ for simplicity).
In particular, this is true for AdS$_5$, as well as the asymptotically AdS$_5$ black hole metric.
In general, applicability of the TsT transformation explained here
and the Kaluza-Klein reduction given in Sec.~\ref{KKreduction} is complementary:
the TsT transformation can be used to obtain solutions with non-constant $h^2$
but our Kaluza-Klein reduction allows us to lift arbitrary solutions of five-dimensional
equations of motion to ten-dimensional solutions.

Notice that it is straightforward to extend the procedure described above to more complicated
black hole geometries, with asymptotic Schr\"odinger symmetry. For example, one can apply the TsT transformation
to the R-charged asymptotically AdS$_5$ black holes constructed in \cite{BuchelLiu}.

\subsection{Conserved charges and thermodynamic properties}
\label{thermodetails}

We will now discuss  some
properties of the black hole metric
\nref{sch:bh} derived in subsection \ref{dipole:bh}.
This can be most simply done by first Kaluza-Klein reducing the metric along the
$\varphi$ and $x^-$ directions to an eight dimensional solution. After this Kaluza-Klein
reduction both the original black three brane metric \nref{blackthree} and the
solution \nref{sch:bh} are related by a symmetry of the eight dimensional gravity theory.
Then, it can be checked explicitly that horizon is not affected by the TsT transformation, and in particular
the Hawking temperature $T$, entropy $S$, and chemical potential $\mu_N$, are unchanged \cite{horowitz}.
We emphasize that it is not a symmetry of the full theory. But since we are computing
quantities just in the gravity theory the results will be the same in both cases.

Let us next address the important issue
of computing conserved charges in backgrounds with asymptotic
Schr\"odinger symmetry such as \nref{metricdip}.
For backgrounds that are asymptotically AdS  there  are several methods
 based on the Fefferman-Graham expansion of the metric near the AdS boundary (see e.g. \cite{Balasubramanian:1999re,Kraus:1999di,de Haro:2000xn,Bianchi:2001kw}).
 In particular, there exist
coordinates such that  any metric which is asymptotically AdS takes  the form
\begin{equation}
\diff s^2_5\, =\, \frac{\diff r^2}{r^2}+r^2\gamma_{ab}\diff x^a \diff x^b
\end{equation}
and $\gamma_{ab}$ has an expansion of the form
\begin{equation}
\gamma_{ab}\, =\,  \gamma^{(0)}_{ab}+\frac{1}{r^2} \gamma^{(2)}_{ab}+\frac{1}{r^4}\gamma^{(4)}_{ab} + h^{(4)}_{ab}\frac{1}{r^4} \log \frac{1}{r^2} +
\cdots
\label{FeffermanGraham}
\end{equation}
Here $\gamma^{(2)}_{ab}$ is  determined in terms of
$\gamma^{(0)}_{ab}$ while $\gamma^{(4)}_{ab}$ captures the leading
deformation with respect to the vacuum. The coefficient $h^{(4)}_{ab}$
is related to the Weyl anomaly  \cite{Henningson:1998gx}
and all other terms are determined recursively in terms of these \cite{de Haro:2000xn}.

This expansion provides good control over the asymptotic behavior of the metric,
allowing  to subtract the infinities consistently.
Then one can add suitable local counterterms and
define the renormalised  boundary energy-momentum tensor as
\begin{equation}
T_{ab}\, =\, \lim_{\epsilon \to 0}\, \frac{1}{8\pi G_N \epsilon^2}\,
\frac{2}{\sqrt{- \gamma}}\frac{\delta}{\delta \gamma^{ab}} \left(S + S_{ct}\right)
\end{equation}
where the integrals are evaluated at a finite distance from the boundary $r =1/\epsilon$.
Notice this is simply
\begin{equation}
T_{ab}\, = \, \gamma^{(4)}_{ab}
\label{Tgamma}
\end{equation}
in the case $\gamma^{(0)}_{ab}=\eta_{ab}$ (then we also have $\gamma^{(2)}_{ab}=h^{(4)}_{ab} =0$). For example, using this method,
 the energy-momentum for the non-extremal
D3-brane metric \eqref{blackthree} can be easily computed and is given by
\begin{equation}
T_{++}\,=\,\frac12\lambda^{-2} r_0^4~,\qquad
T_{--}\,=\,\frac12\lambda^{2} r_0^4~,\qquad
T_{+-}\,=\,\frac14 r_0^4~,\qquad
T_{ij}\,=\,\frac14r_0^4 \delta_{ij}~.
\end{equation}
In general, the conserved charges associated to a Killing vector $\xi^a$ are constructed as
\begin{equation}
Q_\xi \, =\, \int_\Sigma \diff^{3} x \sqrt{\sigma} \, \xi^b T_{ab} u^a
\end{equation}
where $\Sigma$  is a  space-like surface with unit normal vector $u^a$, and is  $\sigma_{ab}$ the induced metric on it. These are easily extracted from the usual  ADM decomposition of the boundary metric, with respect to a chosen time coordinate
\begin{equation}
\diff s^2 \,=\,  -N_\Sigma^2 \diff t^2+\sigma_{ab} (\diff x^a +N_\Sigma^a \diff t)(\diff x^b +N_\Sigma^b \diff t)~.
\end{equation}
Everywhere in the paper we slice at constant $x^+$. This is the natural choice dictated by the
embedding of the Schr\"odinger group into the Poincar\'e group. We can then compute the (non-relativistic) energy, associated to the Killing vector $\de/\de x^+$
\begin{equation}
H \, =\, - P_+ \, =\,  \int_\Sigma \diff^3 x \sqrt{\sigma}(u^+T_{++}+u^- T_{-+})~,
\end{equation}
and the mass (particle number), associated to the Killing vector $\de /\de x^-$
\begin{equation}
\frac{N}{r^-}\, =\, - P_- \, = \, \int_\Sigma  \diff^3 x \sqrt{\sigma}(u^+ T_{+-} + u^- T_{--}) ~.
\end{equation}

There are some obstacles to applying this  program  to  metrics
which asymptote to the
Schr\"odinger background\footnote{More generally, for any value of the exponent $z$ in
(\ref{nrmetric}).}
\bea
\diff s^2 \, =\,  \frac{\diff r^2}{r^2} +
r^2 \left( - \diff x^+\diff x^- +\diff x^i\diff x^i \right)
 -\sigma^2  r^4(\diff x^+)^2~.
\label{schmetric2}
\eea
Firstly, an analogue of the Fefferman-Graham expansion for this kind of asymptotic is not known. In particular, recall
that (\ref{schmetric2}) is not an Einstein metric. Moreover, it is not a priori clear how to define the boundary,
since the leading term is one dimensional.
Secondly,  the full solution is intrinsically ten dimensional, involving both squashing of the
internal geometry, as well as non-trivial RR and NS fields. Application of holographic renormalisation
techniques to ten dimensional geometries is not very well developed (however, see
\cite{Taylor:2001fe,Skenderis:2006uy}). We expect
that the five-dimensional truncations that we derived in this paper will be important
 in formulating a holographic
renormalisation procedure, using for instance the Hamilton-Jacobi approach
\cite{Martelli:2002sp,Papadimitriou:2004ap}.
Of course the whole discussion really makes sense only when $x^-$ is non-compact, otherwise
we cannot trust the metric near the boundary.

We leave a systematic treatment  for future work and instead  circumvent the problem taking advantage
of  the TsT transformation\footnote{The original version of this paragraph had mistakes, we thank
Y. Oz and S. Yankielowicz for pointing them out.}. This transforms the original metric $g_{ab}$ into a new
metric given by   \nref{TsT:metric}.
The proper  Fefferman-Graham expansion of this metric should be such that it coincides with the
corresponding expansion for the original metric. In other words, given the new metric $\tilde g_{ab}$
in \nref{TsT:metric} which was in the string frame,
we can find the original metric $g_{ab}$ as
\be
  g_{ab} = \tilde g_{ab} + \sigma^2 h^2 {  \tilde g_{- a } \tilde g_{-b } \over
  ( 1 - h^2 \sigma^2 \tilde g_{--} ) }
  \ee
  Thus, if someone gives us the metric $\tilde g_{ab}$, we find the metric $g_{ab}$ and
  we perform the usual  Fefferman-Graham expansion
 \eqref{FeffermanGraham},   reading  off the stress tensor
 as in \nref{Tgamma}. The parameters $\sigma $ and $h$ can be read off from the other components
 of the metric and from  the $B$ field.

Then it is natural  to \emph{define} the boundary energy-momentum tensor using $\gamma^{(4)}$ via \eqref{Tgamma}.
Notice that indeed this prescription satisfies the first law of black hole thermodynamics, which is a check that it is correct.
A field theoretical argument  was also given in Sec.~\ref{dipole:bh},
which supports our interpretation here.
It would be worthwhile to
further justify this prescription from the gravity point of view,
and to develop  the holographic renormalisation with this asymptotics.

\section{Details of the consistent truncations}
\label{detailsreduction}

\subsection{Reduction with $m^2=8$}
\label{KKdetails}

We provide the details of the Kaluza-Klein reduction discussed in Sec.~\ref{KKreduction},
leading to \eqref{5d:5daction}.
Let us first use the metric ansatz of the form \begin{equation}
\diff s^2_{10} \,=\,\diff s^2(M)_\text{string}
+\ex^{2U} \diff s^2(B_\text{KE}) + \ex^{2V} (\eta+\cA)^2~.
\end{equation} Note that the five-dimensional part of the metric differs from what is
used in \eqref{5d:ansatz} by a Weyl transformation. $\cA$ is a connection on $M$
which will be set to zero in this subsection. It is included here because we turn it non-zero
in the reduction with $m^2=24$.

The Ricci tensor  has the following components in the flat indices:
\begin{align}
R_{ab}&= R^{(5)}_{ab}-4(\partial_a U \partial_b U+ \nabla_a \partial_b U)
-(\partial_a V\partial_b V+ \nabla_a\partial_b V)-\frac12\ex^{2V}\cF_{ac}\cF_b{}^c~,\label{Ricci1}\\
R_{ij}&=\delta_{ij}(6\ex^{-2U}-2\ex^{-4U+2V}-4\partial_a U\partial^a U-
\partial_a U \partial^a V - \square_5 U)~,\\
R_{\phi\phi}&=4 \ex^{-4U+2V}-4 \partial_a U\partial^a V-\partial_a V \partial^a V
-\square_5 V+\frac14 \ex^{2V}\cF_{ab}\cF^{ab}~,\\
R_{ai}&=R_{i\phi}=0\label{5d:off:ricci}~,\\
R_{a\phi}&=-\frac12 \ex^{-4U-2V}\nabla^b \ex^{4U+3V} \cF_{ba}~,
\label{Ricci2}
\end{align}
where $R^{(5)}_{ab}$ is the Ricci tensor of $\diff s^2_M$ and
the covariant derivatives are with respect to $\diff s^2_M$; $\cF=\diff\cA$ is the curvature
of $\cA$.
For $Y=S^5$ or $Y=T^{1,1}$,
vanishing of $R_{ai}$ and $R_{i\phi}$ immediately follows from the symmetry
of the K\"ahler-Einstein base $\cp^2$ or $\cp^1\times\cp^1$ respectively.
For generic Sasaki-Einsteins one needs to calculate explicitly to see that they vanish.

The field strength $H$ of the two-form $B$ is \begin{equation}
H\,=\,F \wedge \eta -2 A\wedge J~,
\end{equation} so we have
\begin{equation}
H_{ABC}H^{ABC}\,=\, 3 \ex^{-2V} F_{ab}F^{ab} +48 \ex^{-4U} A_a A^a~.
\end{equation}

Plugging these into the 10-dimensional action \eqref{5d:10daction},
we obtain the following 5d action:
\begin{multline}
S\,=\,\frac12\int \diff^5x\sqrt{-g} \ex^{4U+V} \Bigl[ R^{(5)}
+24 \ex^{-2U}-4 \ex^{-4U+2V} - 8 \ex^{-8U-2V} -\frac12\partial_a\Phi\partial^a \Phi \\[2mm]
+12\partial_a U \partial^a U + 8 \partial_a U \partial^a V
-\frac14 \ex^{-\Phi-2V}F_{ab}F^{ab}- 4 \ex^{-\Phi-4U}A_a A^a\Bigr]~.
\end{multline}

By construction, any ansatz of our form \eqref{5d:ansatz}
which is a solution of 10d equations of motion, gives
a 5d metric $\diff s^2(M)$, the scalars $U$, $V$ and $\Phi$ and the gauge field $A$
which solve the equations of motion of the 5d action \eqref{5d:5daction}.
An interesting fact is that the converse is true, i.e. given a solution
to the equations of motion for \eqref{5d:5daction},
the 10d fields constructed along our ansatz
\eqref{5d:ansatz}
automatically solve the 10d equations of motion. In other words,
ours is a consistent  reduction including massive gauge fields and scalars.

Indeed, the $F_5$ is self-dual and closed by construction,
$F_5\wedge H=0$ is also automatic;
the $\Phi$ equation of motion is exactly the same in 5d and in 10d;
moreover it is easily verified that $H$ satisfies $\diff \star H=0$ if
$A$ solves the Proca equation which follows from \eqref{5d:5daction}.

Therefore the only thing to check is the equations of motion of the metric, \nref{5d:metricEOM}.
Let us denote the equations as
\begin{equation}
0=E_{MN}\equiv R_{MN}-(\text{right hand side of \eqref{5d:metricEOM}})~.
\end{equation}
The ``off-diagonal'' parts, \emph{i.e.} $E_{MN}$ with $(M,N)=(a,i)$, $(a,\phi)$, and $(i,\phi)$,
are automatically zero because $R_{MN}$, $\partial_M\Phi \partial_N\Phi$,
$H_{MAB}H_N{}^{AB}$ and $F_{MABCD}F_N{}^{ABCD}$ are all automatically zero.
Moreover, $E_{ij}$ is proportional to $\delta_{ij}$, $E_{ij}=E_B \delta_{ij}$.
So the non-trivial equations are \begin{equation}
E_{ab}\,=\,0~,\quad
E_B\,=\,0~,\quad
E_{\phi\phi}\,=\,0~.
\end{equation}
However, a direct calculation shows that $E_{ab}=0$ is the equations of motion
of the 5d metric $\diff s^2_M$, $E_B=0$ is the one for the scalar $U$, and $E_{\phi\phi}=0$
is the one for the scalar $V$. It is as it should be, because $E_B$ is by definition the variation
of the action with respect to the size of the K\"ahler-Einstein base, which is controlled by $U$,
 etc.
This concludes the proof that the reduction is consistent.

Let us change  the 5d action to the 5d Einstein frame, which is the form presented
in the main part of the paper.
This  can be achieved by
\begin{equation}
\diff s^2(M)_\text{string} \,=\, \ex^{-\frac83U-\frac23V} \diff s^2(M)~.
\end{equation}
Then the action becomes \begin{multline}
S=\frac12\int \diff^5x\sqrt{-g} \Bigl[ R
+24 \ex^{-\frac{14}3U-\frac23V}-4 \ex^{-\frac{20}3U+\frac43V} - 8 \ex^{-\frac{32}{3}U-\frac83V} \\
-\frac12\partial_a\Phi\partial^a \Phi
-\frac{28}{3}\partial_a U \partial^a U -\frac{8}3 \partial_a U \partial^a V-\frac{4}3\partial_a V\partial^a V\\
-\frac14 \ex^{-\Phi+\frac83U-\frac43 V}F_{ab}F^{ab}- 4 \ex^{-\Phi-4U}A_a A^a\Bigr]~.
\end{multline}
Finally one diagonalises the scalar kinetic term by setting
$u=\frac25(U-V)$ and $v=\frac{4}{15}(4U+V)$ to arrive at \eqref{5d:5daction}.
The final result  agrees with the result in \cite{hep-th/0002159} and
in Appendix C of \cite{Benvenuti:2005qb} after setting $\Phi$ and $A$ to zero,
where the Kaluza-Klein reduction
with the metric, $U$ and $V$ was performed. Their scalars $q$, $f$ are related to
ours by a factor of two, $f=u/2$ and $q=v/2$.

\subsection{Reduction with $m^2=24$}
\label{msq24detail}

The ansatz we consider is
\begin{align}
\diff s^2&=\diff s^2(M)_\text{string}+\ex^{2U}\diff s^2(B_\text{KE})+\ex^{2V}(\eta+\cA)^2~,\\
F_5&= 4\ex^{-4U-V}\vol(M)+4\ex^{-4U+V}(\eta+\cA)\wedge \star_5 \bA
+\ex^{-V}\omega\wedge \star_5\bbF \nn\\
&\qquad\qquad +2\omega^2\wedge (\eta+\cA)+2\omega^2 \wedge \bA-\omega
\wedge(\eta+\cA)\wedge \bbF~,
\end{align}
and all other fields vanishing. We denote $\cF=\diff\cA$ and $\bF=\diff\bA$.
We will go to the five-dimensional Einstein frame in the last step.
$F_5=\star F_5$ holds by construction.  $\diff F_5=0$ imposes
\begin{align}
\diff  (\ex^{-4U+V}\star_5 \bA ) &=0~, & (\text{$\eta$ component})~~\nn\\
\diff  ( \ex^{-V}\star_5 \bbF ) &=-8\ex^{-4U+V}\star_5\bA +\cF\wedge\bbF ~, & (\text{$\omega$
component}) \; ~\nn\\
\diff   \bbF&=0~, & (\text{$\omega\wedge\eta$ component})~~\nn\\
\bbF&=\cF+\bF~, & (\text{$\omega\wedge\omega$ component})~.
\end{align}
The components of the Ricci tensor  was tabulated in \eqref{Ricci1},\ldots,\eqref{Ricci2}.
The Einstein equation in ten dimensions is given by $R_{MN}=Q_{MN}$ where
$Q_{MN}\equiv \tfrac{1}{96}F_{MABCD}F_N{}^{ABCD}$, which has the following values in the flat indices:
\begin{eqnarray}
Q_{ab}&=& -4\ex^{-8U-2V}\eta_{ab}\nn\\
&&  +4\ex^{-8U}(2\bA_a \bA_b - \eta_{ab} \bA_c \bA^c)
+\frac14\ex^{-4U-2V} (4\bbF_{ac}\bbF_{b}{}^{c} - \eta_{ab} \bbF_{cd} \bbF^{cd})~ ,\\\
Q_{ij}&=&\delta_{ij}\left(4\ex^{-8U-2V}+4\ex^{-8U}\bA_a\bA^a \right)~,\\
Q_{\phi\phi}&=&4\ex^{-8U-2V}-4\ex^{-8U}\bA_a\bA^a+\frac14\ex^{-4U-2V}\bbF_{ab}
\bbF^{ab}~,\\
Q_{ai}&=&Q_{i\phi}=0~,\\
Q_{a\phi}&=&8\ex^{-8U-V}\bA_a -\frac18\ex^{-4U-2V} \epsilon_{abcde}\bbF^{bc}
\bbF^{de}~.
\end{eqnarray}
Then $R_{a\phi}=Q_{a\phi}$ gives
 \begin{align}
\diff (\ex^{4U+3V}\star_5 \cF ) &=16\ex^{-4U+V}\star_5 \bA+\bbF\wedge \bbF~,
\end{align} $R_{ab}=Q_{ab}$ is \begin{align}
R^{(5)}_{ab}&=4(\partial_a U \partial_b U+ \nabla_a \partial_b U) +
(\partial_a V\partial_b V+ \nabla_a\partial_b V)-4\ex^{-8U-2V} \eta_{ab}\nn\\
&\quad +\frac12\ex^{2V}\cF_{ac}\cF_b{}^c+4\ex^{-8U}(2\bA_a \bA_b - \delta_{ab}
\bA_c \bA^c)
+\frac14\ex^{-4U-2V} (4\bbF_{ac}\bbF_{b}{}^{c} - \delta_{ab} \bbF_{cd} \bbF^{cd})~,
\end{align} and finally $R_{ij}=Q_{ij}$ and $R_{\phi\phi}=Q_{\phi\phi}$ give respectively
\begin{align}
\square_5 U&=6\ex^{-2U}-2\ex^{-4U+2V}-4\partial_a U\partial^a U-
\partial_a U \partial^a V -4\ex^{-8U-2V}-4\ex^{-8U}\bA_a\bA^a~,\\
\square_5 V&=4 \ex^{-4U+2V}-4 \partial_a U\partial^a V-\partial_a V \partial^a V
+\frac14 \ex^{2V}\cF_{ab}\cF^{ab} -4\ex^{-8U-2V}\nn\\
&\quad +4 \ex^{-8U}\bA_a \bA^a -\frac14\ex^{-4U-2V}\bbF_{ab}\bbF^{ab}~.
\end{align}
This set of equations comes from the 5d action
\begin{multline}
S=\frac12\int \diff^5x\sqrt{-g} \ex^{4U+V} \Bigl[ R^{(5)}
+24 \ex^{-2U}-4 \ex^{-4U+2V} - 8 \ex^{-8U-2V}
+12\partial_a U \partial^a U + 8 \partial_a U \partial^a V\\
-\frac14 \ex^{2V}\cF_{ab}\cF^{ab}
- \frac12  \ex^{-4U-2V} \bbF_{ab} \bbF^{ab} -8\ex^{-8U} \bA_a\bA^a \Bigr]
+\frac12\int \cA\wedge \bbF \wedge \bbF~.
\end{multline} To check the equations of motion,
the following formula is useful:
 \begin{equation}
\frac1{\sqrt{-g} }\frac{\delta}{\delta g^{ab}}\int \diff^n x  \sqrt{-g}X R
\,=\, X(R_{ab}-\frac12g_{ab}R)-\nabla_a\partial_b X + g_{ab}\square X~,
\end{equation}
where $X$ is a scalar field.
Changing to the Einstein frame in five dimensions,
we obtain the action \eqref{5daction:m2=24}.


\section{Supersymmetry analysis of the Schr\"odinger vacuum}
\label{schsusy}

Here we study the supersymmetry preserved by the background \eqref{metricdip}.
Supersymmetric extensions of the non-relativistic conformal groups were studied in \cite{ABS,Duval:1993hs,Leblanc:1992wu,Sakaguchi:2008rx,Sakaguchi:2008ku}.
Hereafter we denote  the generators of  the supertranslation and the special superconformal transformations of the relativistic superconformal algebra  by $Q_\pm$, $S_\pm$, respectively. The subscripts $\pm$ shows the charge under the boost, so we have
for example $\{Q_+,Q_+\}\propto \tilde P_+$.
We find that the solution does not preserve any
supersymmetry\footnote{We thank the authors of \cite{HRR} for pointing out the error
in our statement in the first version of this paper.}.

Let us first recall the supersymmetry variations of the fermions in the theory.
We follow the notations in \cite{Gauntlett:2005ww}. In particular,
we combine two Majorana-Weyl spinors $\epsilon^{1,2}$ of type IIB supergravity
into a complex Weyl spinor $\epsilon=\epsilon^1+i \epsilon^2$
and $\epsilon^c$ is its complex conjugate, $\epsilon^c=\epsilon^1-i\epsilon^2$.
Then the gravitino transformation law in the Einstein frame is
\begin{equation}
\delta \psi_M = D_M \epsilon -\frac{i}{96}\ex^{-\Phi/2}\left(
\Gamma_M{}^{ABC}H_{ABC}-9\Gamma^{AB}H_{MAB}
\right)\epsilon^c+
 \frac{i}{192}\Gamma^{ABCD}F_{MABCD} \epsilon~,
\end{equation}
and that for the dilatino is
\begin{equation}
\delta \lambda \, =\, -\frac{1}{24}\ex^{-\Phi/2}\Gamma^{ABC}H_{ABC}\epsilon~,
\end{equation}
where we have set the RR-axion and the RR 2-form to zero since they do not appear
in our ans\"atze.

Let us determine first  the  supersymmetry preserved by the
AdS$_5\times S^5$ compactification of type IIB string theory,
where the $x^-$ direction is compactified to break the relativistic conformal symmetry
to the Schr\"odinger symmetry.
The gravitino variation is
\begin{equation}
\delta \psi_\mu \,=\, \partial_\mu \epsilon - \frac 12 \Gamma_\mu M \epsilon ~, \qquad M \, \equiv \,\left( \Gamma_4 -
\frac{i}{4} F_{01234} \Gamma^{01234}\right)
\label{ads-susy}
\end{equation}
for $\mu=0,1,2,3$.
Recalling $F_{01234}=4$ in our convention, the operator $M$ above
vanishes in the sector where $i\Gamma^{0123}=1$,
which fixes the  chirality of $\epsilon$,
where $\epsilon$ is independent of $x^{+,-,2,3}$.
Let us denote this ``$Q$ sector'' temporarily.

The commutator of two supersymmetry transformation gives
\begin{equation}
[\delta_\epsilon,\delta_{\epsilon'}]\propto
(\epsilon \gamma^M \epsilon') \partial_M ;
\end{equation}
so the commutator of two $\epsilon$'s in the sector $Q$
gives a diffeomorphism which is independent of $x^{+,-,2,3}$.
They correspond to the translations $\tilde P^\mu$
which in turn means the sector $Q$ generates supertranslations $Q_\pm$,
and other spinors with opposite chirality $i\Gamma^{0123}=-1$ correspond
to the special superconformal generators $S_\pm$.

Let us compactify the $x^-$ direction, $x^-\sim x^-+ 2\pi r^-$.
We can choose boundary conditions for the fermion, but under the most naive
one the preserved supersymmetries are independent of $x^-$.
Then \eqref{ads-susy} requires
\begin{equation}
\Gamma^+(1-i\Gamma^{0123})\epsilon\,=\, 0~.
\end{equation} Therefore the preserved supersymmetry generators are
$Q_\pm$ and $S_-$, and the superalgebra of the background preserves $3/4$ of
the original supersymmetry.

Now we will show that the  Schr\"odinger
vacuum
\eqref{sch:metric} does not preserve supersymmetry.
We will show that a Killing spinor does not exist.
Using the value of the $H$ field, we see that setting to zero the dilatino variation
we get the condition
\begin{equation}
0=\delta \lambda \,=\,  -\frac{\sigma}2 \Gamma^+ \slash { \omega }_C \epsilon~.
\end{equation}
This imposes $\Gamma^+\epsilon=0$ because
$\slash {\omega}_C$ is clearly  invertible.
Then, consider the equation that arises from the $\delta\psi_+$ part of the gravitino variation
\begin{equation}
0=\delta\psi_+ \,=\, \frac1r\partial_+ \epsilon- \frac 12 \Gamma_+ M\epsilon +\frac{3i\sigma}8
\slash{\omega}_C \epsilon^c .
\end{equation}
The compatibility of $\Gamma^+\epsilon=0$
and the evolution along $x^+$  imposes $i\Gamma^{0123}\epsilon=\epsilon$.
It implies $i\Gamma^{0123}\epsilon^c=-\epsilon^c$, which means
that we cannot impose $i\Gamma^{0123}\epsilon(x^+)=\epsilon(x^+)$ for all $x^+$.
Therefore there is no supersymmetry preserved by this background.

\section{Note Added: Scaling behavior and the gravity background}\footnote{This appendix
is a note added in Nov, 2008.}
\label{powerlaw}
Invariance under the  Schr\"odinger group imposes a certain constraint on the thermodynamic quantities,
e.g.~the energy density $H/V_d$ of a $d$-dimensional system.
Recall that the dilatation $x\to kx$, $t\to k^2 t$ has the  effect \begin{equation}
T\to k^{-2} T,\qquad
\frac{H}{V_d} \to k^{-2-d} \frac{H}{V_d},\qquad
\mu \to k^{-2} \mu.
\end{equation} Then it follows that the energy density as a function of $\mu$ and $T$ has the form
\begin{equation}
\frac{H}{V_d}=T^{1+\frac{d}2} g(\frac\mu T) \label{gg}
\end{equation} where $g(x)$ is a function.
When the result of Sec.~\ref{DLCQofAdS} is expressed in this form, we find that \begin{equation}
g(x) \propto x^{-2}
\end{equation} which shows a simple scaling behavior, unlike the corresponding function for a free non-relativistic gas of bosons or fermions,
as noticed in \cite{Kovtun}. This power-law behavior of the function $g(x)$ is a direct consequence
of the fact that the gravity dual  is a light-like compactification, as will be explained below.
In fact, before the light like compactification the gravity theory has a boost symmetry in the 
$x^\pm$ directions. The compactification breaks this symmetry but in a very controlled way which
leads to the above behavior for the function $g(x)$. 

Recall that the dilatation of the non-relativistic system arose as the transformation
on the gravity background,
\begin{equation}
x^+\to k^2 x^+,\qquad x^i \to  k x^i, \qquad r\to k^{-1} x^i,
\end{equation} which in particular is an isometry of the zero-temperature solution.
Now instead let us boost the $x^\pm$ plane \begin{equation}
x^+\to s x^+, \qquad x^- \to s^{-1} x^-,
\end{equation}  which changes $r^-$, the compactification radius of $x^-$, to $s^{-1}r^-$.
This establishes a correspondence between the theory with parameter $r^-$ to the theory 
with parameter $ s^{-1} r^-$. On the other hand, when we change the value of $r^-$ we change
the gravity results in a very simple way. The reason is that many quantities have
a simple dependence on $r^-$ since the geometry is basically independent of $r^-$ and some
quantities, such as the energy, are extensive in $r^-$. We should emphasize that this simple
dependence on $r^-$ is a property of the classical gravity solutions we are considering, but quantum 
corrections will not have this simple behavior. For example, the Casimir energy could have
a more complicated dependence on $r^-$. However, if we simply focus on the leading classical
gravity approximation we can figure out precisely how the results change
  under the combined effects of the boost and the $r^-$ rescaling. 
This operation makes the following change in the thermodynamic variables: \begin{equation}
T\to s^{-1}T,\qquad
\frac{H}{V_d} \to \frac{H}{V_d},\qquad
\mu \to s^{-2}\mu.
\end{equation}
Therefore the energy density can be written as\begin{equation}
\frac{H}{V_d}= f(\frac{\mu}{T^2}) \label{ff}
\end{equation} where $f(x)$ is an arbitrary function.
The combination of \eqref{gg} and \eqref{ff} then implies \begin{equation}
\frac{H}{V_d} \propto \frac{T^{2+d}}{\mu ^{1+\frac d2}},
\end{equation} which reproduces the result of an explicit calculation in \cite{Kovtun}.
We can similarly state how the entropy density and number particle density scale
\be
 { S \over V_d} \propto  T^{d+1} \mu^{ -1 - {d \over 2}}  ~,~~~~~~~~~~~~~~~~
 { N \over V_d} \propto T^{d+2} \mu^{-2 -{ d \over 2} }.
 \ee
 
The derivation above shows that this power-law behavior of thermodynamic quantities
is a universal property of a system with non-relativistic symmetry which arises as a DLCQ
of a relativistic conformal theory with dual gravity description.
For example, the DLCQ of the dipole theory discussed in Sec.~\ref{dipole:bh}
also shows this behavior. Once quantum corrections to the gravity results are computed, we 
could get different answers. Here we also assumed that the gravity saddle point is translational
invariant along $x^\pm$, if that were not the case, we would get different scaling properties.


\begin{thebibliography}{99}
\bibitem{son}
 D.~T.~Son,
 ``Toward an AdS/cold atoms correspondence: a geometric realization of the
 Schroedinger symmetry,''
 {\slshape Phys.\ Rev.\  D\ } {\bf 78}, 046003 (2008)
 [arXiv:0804.3972 [hep-th]].

 \bibitem{mcgreevy}
  K.~Balasubramanian and J.~McGreevy,
 ``Gravity duals for non-relativistic CFTs,''
{\slshape Phys.\ Rev.\ Lett.\  } {\bf 101}, 061601 (2008)
 [arXiv:0804.4053 [hep-th]].

 
  \bibitem{goldberger}
  W.~D.~Goldberger,
  ``AdS/CFT duality for non-relativistic field theory,''
  arXiv:0806.2867 [hep-th].

 \bibitem{barbonfuertes}
 
 
 J.~L.~B.~Barbon and C.~A.~Fuertes,
 ``On the spectrum of nonrelativistic AdS/CFT,''
{ \slshape JHEP\ } {\bf 0809}, 030 (2008)
 [arXiv:0806.3244 [hep-th]].

 
  \bibitem{Horvathy:2003rb}
  P.~A.~Horv\'athy,
  ``Nonrelativistic conformal structures,''
  arXiv:math-ph/0305054.

 \bibitem{Duval:1990hj}
  C.~Duval, G.~W.~Gibbons and P.~Horv\'athy,
  ``Celestial Mechanics, Conformal Structures, and Gravitational Waves,''
 {\slshape   Phys.\ Rev.\  D }{\bf 43} (1991) 3907
  [arXiv:hep-th/0512188].

 \bibitem{nishidason}
  Y.~Nishida and D.~T.~Son,
  ``Nonrelativistic conformal field theories,''
 {\slshape   Phys.\ Rev.\  D }{\bf 76} (2007) 086004
  [arXiv:0706.3746 [hep-th]].

 \bibitem{ABS}
  O.~Aharony, M.~Berkooz and N.~Seiberg,
  ``Light-cone description of (2,0) superconformal theories in six dimensions,''
 {\slshape   Adv.\ Theor.\ Math.\ Phys.\  }{\bf 2} (1998) 119
  [arXiv:hep-th/9712117].

 \bibitem{stromingerlectures}
  R.~Britto-Pacumio, J.~Michelson, A.~Strominger and A.~Volovich,
  ``Lectures on superconformal quantum mechanics and multi-black hole moduli spaces,''
  arXiv:hep-th/9911066.

 \bibitem{dipolelightlike}
  M.~Alishahiha and O.~J.~Ganor,
  ``Twisted backgrounds, pp-waves and nonlocal field theories,''
 {\slshape   JHEP }{\bf 0303} (2003) 006
  [arXiv:hep-th/0301080].

 \bibitem{dipolegravity}
  A.~Bergman, K.~Dasgupta, O.~J.~Ganor, J.~L.~Karczmarek and G.~Rajesh,
  ``Nonlocal field theories and their gravity duals,''
 {\slshape   Phys.\ Rev.\  D }{\bf 65} (2002) 066005
  [arXiv:hep-th/0103090].

 \bibitem{filk}
  T.~Filk,
  ``Divergencies in a field theory on quantum space,''
 {\slshape   Phys.\ Lett.\  B }{\bf 376} (1996) 53.

 \bibitem{ABM}
  A.~Adams, K.~Balasubramanian and J.~McGreevy,
  ``Hot Spacetimes for Cold Atoms,''
  arXiv:0807.1111 [hep-th].

 \bibitem{HRR}
  C.~P.~Herzog, M.~Rangamani and S.~F.~Ross,
  ``Heating up Galilean holography,''
  arXiv:0807.1099 [hep-th].

 \bibitem{maskawa}
T. Maskawa and K. Yamawaki, {\slshape Prog. Theor. Phys.\ } \textbf{56} (1976) 270;
A. Casher, {\slshape Phys. Rev. D\ } \textbf{14} (1997) 452;
R. Giles and C. B. Thorn, {\slshape Phys. Rev. D\ } \textbf{16} (1977) 366;
C. B. Thorn, {\slshape Phys. Rev. D\ } \textbf{19} (1979) 639;
H. C. Pauli and S. J. Brodsky, {\slshape Phys. Rev.  D\ } \textbf{32} (1985) 1993, 2001.

\bibitem{shjp}
  S.~Hellerman and J.~Polchinski,
  ``Compactification in the lightlike limit,''
 {\slshape   Phys.\ Rev.\  D }{\bf 59} (1999) 125002
  [arXiv:hep-th/9711037].

 \bibitem{ganorsethi}
  O.~J.~Ganor and S.~Sethi,
  ``New perspectives on Yang-Mills theories with sixteen supersymmetries,''
 {\slshape   JHEP }{\bf 9801} (1998) 007
  [arXiv:hep-th/9712071].

 \bibitem{kapustinsethi}
  A.~Kapustin and S.~Sethi,
  ``The Higgs branch of impurity theories,''
 {\slshape   Adv.\ Theor.\ Math.\ Phys.\  }{\bf 2} (1998) 571
  [arXiv:hep-th/9804027].

 \bibitem{stromingeradstwo}
  A.~Strominger,
  ``AdS$_2$ quantum gravity and string theory,''
 {\slshape   JHEP }{\bf 9901} (1999) 007
  [arXiv:hep-th/9809027].

 \bibitem{dinerajaraman}
  M.~Dine and A.~Rajaraman,
  ``Multigraviton scattering in the matrix model,''
 {\slshape   Phys.\ Lett.\  B }{\bf 425} (1998) 77
  [arXiv:hep-th/9710174].

 \bibitem{flattopp}
  C.~Duval, P.~A.~Horv\'athy and L.~Palla,
  ``Conformal properties of Chern-Simons vortices in external fields,''
 {\slshape   Phys.\ Rev.\  D }{\bf 50} (1994) 6658
  [arXiv:hep-ph/9405229].

 \bibitem{dipoleone}
  A.~Bergman and O.~J.~Ganor,
  ``Dipoles, twists and noncommutative gauge theory,''
 {\slshape   JHEP }{\bf 0010} (2000) 018
  [arXiv:hep-th/0008030].

 \bibitem{dipoletwo}
  K.~Dasgupta, O.~J.~Ganor and G.~Rajesh,
  ``Vector deformations of ${\mathcal{N}}\!=4$ super-Yang-Mills theory, pinned branes, and arched strings,''
 {\slshape   JHEP }{\bf 0104} (2001) 034
  [arXiv:hep-th/0010072].

 \bibitem{SUSSKIND}
  L.~Susskind,
  ``Another conjecture about M(atrix) theory,''
  arXiv:hep-th/9704080.

 \bibitem{BFSS}
  T.~Banks, W.~Fischler, S.~H.~Shenker and L.~Susskind,
  ``M theory as a matrix model: A conjecture,''
 {\slshape   Phys.\ Rev.\  D }{\bf 55} (1997) 5112
  [arXiv:hep-th/9610043].

 \bibitem{Seibergmm}
  N.~Seiberg,
  ``Why is the matrix model correct?,''
 {\slshape   Phys.\ Rev.\ Lett.\  }{\bf 79} (1997) 3577
  [arXiv:hep-th/9710009].

 \bibitem{IIBDLCQ}
  T.~Banks and N.~Seiberg,
  ``Strings from matrices,''
 {\slshape   Nucl.\ Phys.\  B }{\bf 497} (1997) 41
  [arXiv:hep-th/9702187].

 \bibitem{Hawking:1998kw}
  S.~W.~Hawking, C.~J.~Hunter and M.~Taylor,
  ``Rotation and the AdS/CFT correspondence,''
 {\slshape   Phys.\ Rev.\  D }{\bf 59} (1999) 064005
  [arXiv:hep-th/9811056].

 \bibitem{AdSKerrFirstLaw}
  G.~W.~Gibbons, M.~J.~Perry and C.~N.~Pope,
  ``The first law of thermodynamics for Kerr -- anti-de Sitter black holes,''
 {\slshape   Class.\ Quant.\ Grav.\  }{\bf 22} (2005) 1503
  [arXiv:hep-th/0408217].

 \bibitem{Bhattacharyya:2007vs}
S.~Bhattacharyya, S.~Lahiri, R.~Loganayagam and S.~Minwalla,
 ``Large rotating AdS black holes from fluid mechanics,''
{\slshape JHEP\ } {\bf 0809}, 054 (2008)
 [arXiv:0708.1770 [hep-th]].


 \bibitem{HP}
  S.~W.~Hawking and D.~N.~Page,
  ``Thermodynamics of Black Holes in Anti-de~Sitter Space,''
 {\slshape   Commun.\ Math.\ Phys.\  }{\bf 87} (1983) 577.

 \bibitem{wittenthermal}
  E.~Witten,
  ``Anti-de Sitter space, thermal phase transition, and confinement in gauge theories,''
 {\slshape   Adv.\ Theor.\ Math.\ Phys.\  }{\bf 2} (1998) 505
  [arXiv:hep-th/9803131].

 \bibitem{Bigatti}
  D.~Bigatti and L.~Susskind,
  ``Magnetic fields, branes and noncommutative geometry,''
 {\slshape   Phys.\ Rev.\  D }{\bf 62} (2000) 066004
  [arXiv:hep-th/9908056].

 \bibitem{Martelli:2008cm}
  D.~Martelli and J.~Sparks,
  ``Symmetry-breaking vacua and baryon condensates in AdS/CFT,''
  arXiv:0804.3999 [hep-th].

 \bibitem{hep-th/0002159}
  I.~R.~Klebanov and A.~A.~Tseytlin,
  ``Gravity duals of supersymmetric $SU(N)$ $\times$ $SU(N+M)$ gauge theories,''
 {\slshape   Nucl.\ Phys.\  B }{\bf 578} (2000) 123
  [arXiv:hep-th/0002159].

 \bibitem{de Haro:2000xn}
  S.~de Haro, S.~N.~Solodukhin and K.~Skenderis,
  ``Holographic reconstruction of spacetime and renormalization in the AdS/CFT correspondence,''
 {\slshape   Commun.\ Math.\ Phys.\  }{\bf 217} (2001) 595
  [arXiv:hep-th/0002230].

 \bibitem{Campos:2000yu}
  V.~L.~Campos, G.~Ferretti, H.~Larsson, D.~Martelli and B.~E.~W.~Nilsson,
  ``A study of holographic renormalization group flows in d = 6 and d = 3,''
 {\slshape   JHEP }{\bf 0006} (2000) 023
  [arXiv:hep-th/0003151].

 \bibitem{Martelli:2002sp}
  D.~Martelli and W.~M\"uck,
  ``Holographic renormalization and Ward identities with the Hamilton-Jacobi method,''
 {\slshape   Nucl.\ Phys.\  B }{\bf 654} (2003) 248
  [arXiv:hep-th/0205061].

 \bibitem{Papadimitriou:2004ap}
  I.~Papadimitriou and K.~Skenderis,
  ``AdS / CFT correspondence and geometry,''
  arXiv:hep-th/0404176.

 \bibitem{kim}
  H.~J.~Kim, L.~J.~Romans and P.~van Nieuwenhuizen,
  ``The Mass Spectrum of Chiral ${\mathcal{N}}\!=2$ D=10 Supergravity on $ S^5$,''
 {\slshape   Phys.\ Rev.\  D }{\bf 32} (1985) 389.

 \bibitem{Benvenuti:2006xg}
  S.~Benvenuti, L.~A.~Pando Zayas and Y.~Tachikawa,
  ``Triangle anomalies from Einstein manifolds,''
 {\slshape   Adv.\ Theor.\ Math.\ Phys.\  }{\bf 10} (2006) 395
  [arXiv:hep-th/0601054].

 \bibitem{BuchelLiu}
  A.~Buchel and J.~T.~Liu,
  ``Gauged supergravity from type IIB string theory on $Y^{p,q}$ manifolds,''
 {\slshape   Nucl.\ Phys.\  B }{\bf 771} (2007) 93
  [arXiv:hep-th/0608002].

 \bibitem{Gauntlett:2007ma}
  J.~P.~Gauntlett and O.~Varela,
  ``Consistent Kaluza-Klein Reductions for General Supersymmetric AdS Solutions,''
 {\slshape   Phys.\ Rev.\  D }{\bf 76} (2007) 126007
  [arXiv:0707.2315 [hep-th]].

 \bibitem{Itzhaki}
  N.~Itzhaki, J.~M.~Maldacena, J.~Sonnenschein and S.~Yankielowicz,
  ``Supergravity and the large $N$ limit of theories with sixteen supercharges,''
 {\slshape   Phys.\ Rev.\  D }{\bf 58} (1998) 046004
  [arXiv:hep-th/9802042].

 \bibitem{Anagnostopoulos:2007fw}
  K.~N.~Anagnostopoulos, M.~Hanada, J.~Nishimura and S.~Takeuchi,
  ``Monte Carlo studies of supersymmetric matrix quantum mechanics with sixteen supercharges at finite temperature,''
 {\slshape   Phys.\ Rev.\ Lett.\  }{\bf 100} (2008) 021601
  [arXiv:0707.4454 [hep-th]].

 \bibitem{TW}

 S.~Catterall and T.~Wiseman,
 ``Black hole thermodynamics from simulations of lattice Yang-Mills theory,''
{\slshape  Phys.\ Rev.\  D\ } {\bf 78}, 041502 (2008)
 [arXiv:0803.4273 [hep-th]].

 \bibitem{bulk}
  D.~T.~Son,
  ``Vanishing bulk viscosities and conformal invariance of unitary Fermi gas,''
 {\slshape   Phys.\ Rev.\ Lett.\  }{\bf 98} (2007) 020604
  [arXiv:cond-mat/0511721].

 \bibitem{pss}
  G.~Policastro, D.~T.~Son and A.~O.~Starinets,
  ``The shear viscosity of strongly coupled ${\mathcal{N}}\!=4$ supersymmetric Yang-Mills plasma,''
 {\slshape   Phys.\ Rev.\ Lett.\  }{\bf 87} (2001) 081601
  [arXiv:hep-th/0104066].

 \bibitem{pss2}
  G.~Policastro, D.~T.~Son and A.~O.~Starinets,
  ``From AdS/CFT correspondence to hydrodynamics,''
 {\slshape   JHEP }{\bf 0209} (2002) 043
  [arXiv:hep-th/0205052].

 \bibitem{DFF}
  V.~de Alfaro, S.~Fubini and G.~Furlan,
  ``Conformal Invariance in Quantum Mechanics,''
 {\slshape   Nuovo Cim.\  A }{\bf 34} (1976) 569.

 \bibitem{kerrds}
  G.~W.~Gibbons, H.~L\"u, D.~N.~Page and C.~N.~Pope,
  ``The general Kerr-de Sitter metrics in all dimensions,''
 {\slshape   J.\ Geom.\ Phys.\  }{\bf 53} (2005) 49
  [arXiv:hep-th/0404008].

 \bibitem{Chong:2005hr}
  Z.~W.~Chong, M.~Cveti\v c, H.~L\"u and C.~N.~Pope,
  ``General non-extremal rotating black holes in minimal five-dimensional gauged supergravity,''
 {\slshape   Phys.\ Rev.\ Lett.\  }{\bf 95} (2005) 161301
  [arXiv:hep-th/0506029].

 \bibitem{LLM}
  O.~Lunin and J.~M.~Maldacena,
  ``Deforming field theories with U(1) $\times$ U(1) global symmetry and their gravity duals,''
 {\slshape   JHEP }{\bf 0505} (2005) 033
  [arXiv:hep-th/0502086].

 \bibitem{horowitz}
  G.~T.~Horowitz and D.~L.~Welch,
  ``Duality invariance of the Hawking temperature and entropy,''
 {\slshape   Phys.\ Rev.\  D }{\bf 49} (1994) 590
  [arXiv:hep-th/9308077].

 \bibitem{Balasubramanian:1999re}
  V.~Balasubramanian and P.~Kraus,
  ``A stress tensor for anti-de Sitter gravity,''
 {\slshape   Commun.\ Math.\ Phys.\  }{\bf 208} (1999) 413
  [arXiv:hep-th/9902121].

 \bibitem{Kraus:1999di}
  P.~Kraus, F.~Larsen and R.~Siebelink,
  ``The gravitational action in asymptotically AdS and flat spacetimes,''
 {\slshape   Nucl.\ Phys.\  B }{\bf 563} (1999) 259
  [arXiv:hep-th/9906127].

 \bibitem{Bianchi:2001kw}
  M.~Bianchi, D.~Z.~Freedman and K.~Skenderis,
  ``Holographic renormalization,''
 {\slshape   Nucl.\ Phys.\  B }{\bf 631} (2002) 159
  [arXiv:hep-th/0112119].

 \bibitem{Henningson:1998gx}
  M.~Henningson and K.~Skenderis,
  ``The holographic Weyl anomaly,''
 {\slshape   JHEP }{\bf 9807} (1998) 023
  [arXiv:hep-th/9806087].

 \bibitem{Taylor:2001fe}
  M.~Taylor,
  ``Higher-dimensional formulation of counterterms,''
  arXiv:hep-th/0110142.

 \bibitem{Skenderis:2006uy}
  K.~Skenderis and M.~Taylor,
  ``Kaluza-Klein holography,''
 {\slshape   JHEP }{\bf 0605} (2006) 057
  [arXiv:hep-th/0603016].

 \bibitem{Benvenuti:2005qb}
  S.~Benvenuti, M.~Mahato, L.~A.~Pando Zayas and Y.~Tachikawa,
  ``The gauge / gravity theory of blown up four cycles,''
  arXiv:hep-th/0512061.

 \bibitem{Duval:1993hs}
  C.~Duval and P.~A.~Horv\'athy,
  ``On Schr\"odinger superalgebras,''
 {\slshape   J.\ Math.\ Phys.\  }{\bf 35} (1994) 2516
  [arXiv:hep-th/0508079].

 \bibitem{Leblanc:1992wu}
  M.~Leblanc, G.~Lozano and H.~Min,
  ``Extended superconformal Galilean symmetry in Chern-Simons matter systems,''
 {\slshape   Annals Phys.\  }{\bf 219} (1992) 328
  [arXiv:hep-th/9206039].

 \bibitem{Sakaguchi:2008rx}
  M.~Sakaguchi and K.~Yoshida,
  ``Super Schr\"odinger algebra in AdS/CFT''
{\slshape  J. Math. Phys.\ } {\bf 49} (2008) 102302 
[arXiv:0805.2661 [hep-th]].

 \bibitem{Sakaguchi:2008ku}
 M.~Sakaguchi and K.~Yoshida,
 ``More super Schrodinger algebras from psu(2,2|4),''
{\slshape  JHEP\ } {\bf 0808}, 049 (2008)
 [arXiv:0806.3612 [hep-th]].
 
  \bibitem{Gauntlett:2005ww}
  J.~P.~Gauntlett, D.~Martelli, J.~Sparks and D.~Waldram,
  ``Supersymmetric $\mathrm{AdS}_5$ solutions of type IIB supergravity,''
 {\slshape   Class.\ Quant.\ Grav.\  }{\bf 23} (2006) 4693
  [arXiv:hep-th/0510125].

\bibitem{Kovtun}
  P.~Kovtun and D.~Nickel,
  ``Black holes and non-relativistic quantum systems,''
  arXiv:0809.2020 [hep-th].
\end{thebibliography}
\end{document}